\documentclass[
prl,
twocolumn,
superscriptaddress,
nofootinbib
]{revtex4-1}
\pdfoutput=1
\usepackage[letterpaper,top=1.7cm,bottom=2.54cm,left=1.85cm,right=1.85cm]{geometry}
\usepackage{amssymb,amsmath,amsthm,graphicx,mathptmx}
\usepackage{subfigure}
\usepackage{graphics, color}
\usepackage{latexsym}
\usepackage{bm}
\usepackage{epsfig}
\usepackage{multirow,tabularx}
\usepackage[colorlinks=true,linktocpage=true,linkcolor=blue,citecolor=blue]{hyperref}
\usepackage[abs]{overpic}

\usepackage{fancyhdr} 
\fancypagestyle{plain}

\newcommand{\PRLsection}[1]{\emph{#1.---}}

\newcommand{\TF}{\mathrm{TF}}  %TF
\newcolumntype{Y}{>{\centering\arraybackslash}X}

\begin{document}
\title{\large End Point of Black Ring Instabilities and the Weak Cosmic Censorship Conjecture}
\author{Pau Figueras}%
\email{p.figueras@qmul.ac.uk}
\affiliation{\small \mbox{School of Mathematical Sciences, Queen Mary University of London, Mile End Road, London E1 4NS, United Kingdom}}
 \altaffiliation[Also at ]{DAMTP, Centre for Mathematical Sciences, Wilberforce Road, Cambridge CB3 0WA, United Kingdom}
\author{Markus Kunesch}%
 \email{m.kunesch@damtp.cam.ac.uk}
\author{Saran Tunyasuvunakool}%
 \email{s.tunyasuvunakool@damtp.cam.ac.uk} 
\affiliation{\small \mbox{Department of Applied Mathematics and Theoretical Physics (DAMTP), Centre for Mathematical Sciences, University of Cambridge,} \mbox{Wilberforce Road, Cambridge CB3 0WA, United Kingdom}}

%\date{}

\begin{abstract}
\small
We produce the first concrete evidence that violation of the weak cosmic censorship conjecture can occur in asymptotically flat spaces of five dimensions by numerically evolving perturbed black rings. For certain thin rings, we identify a new, elastic-type instability dominating the evolution, causing the system to settle to a spherical black hole. However, for sufficiently thin rings the Gregory-Laflamme mode is dominant, and the instability unfolds similarly to that of black strings, where the horizon develops a structure of bulges connected by necks which become ever thinner over time.
\end{abstract}


\maketitle
\thispagestyle{fancy}

%-------------------------------------------------------
% Introduction
%-------------------------------------------------------
\PRLsection{Introduction}%
Black holes are amongst the most important solutions of the Einstein equations. Despite their simplicity, they capture some of the most fundamental aspects of the theory. The black holes of general relativity also play a key role in astrophysics, in particular as a description of the compact dark objects at the center of galaxies. This relies on the assumption that they are nonlinearly stable to small perturbations. Although the full nonlinear stability of the Kerr solution \cite{Kerr:1963ud} has not been rigorously proven, there is good evidence that it is indeed stable \cite{Whiting:1988vc,Dafermos:2010hb,Dafermos:2014cua,Dafermos:2014jwa}. 

The situation is markedly different in higher dimensions, where black holes can be dynamically unstable to gravitational perturbations. This was first shown by Gregory and Laflamme \cite{Gregory:1993vy} in the case of black strings and black $p$-branes. Their result was later generalized to boosted black strings \cite{Hovdebo:2006jy}. In a remarkable paper \cite{Lehner:2010pn}, Lehner and Pretorius used numerical relativity techniques to study the nonlinear evolution of the Gregory-Laflamme (GL) instability of the five-dimensional black string. They found that the instability unfolds in a self-similar process which gives rise to a sequence of black hole ``bulges'' connected by black strings which become ever thinner over time. Furthermore, they provided convincing evidence that this process would lead to these thin strings completely pinching off within finite time. This result was interpreted as evidence for a violation of the weak cosmic censorship conjecture (WCC) \cite{Penrose:1969pc,Christodoulou:1999ve} in spacetimes with compact extra dimensions.

Another novel aspect of higher dimensional black hole physics is that horizons can have nonspherical topologies, even in asymptotically flat spaces. The five-dimensional black ring of Emparan and Reall \cite{Emparan:2001wn,Emparan:2006mm} is the first example. This is a stationary solution of the vacuum Einstein equations with horizon topology $S^1\times S^2$. The $S^1$ of the ring is a contractible circle that is stabilized by the centrifugal force provided by the rotation. In terms of the standard dimensionless ``thickness" parameter $\nu$ \cite{Emparan:2006mm}, black rings can be classified as either ``thin" $(0<\nu<0.5)$ or ``fat" $(0.5<\nu<1)$. This thickness parameter describes the relative sizes between the $S^1$ and the $S^2$ of the ring. Fat rings are known to be unstable under radial perturbations \cite{Elvang:2006dd,Figueras:2011he}. Very recently, thin rings have been shown to be linearly subject to a GL-like instability \cite{Santos:2015iua,Tanabe:2015hda}. Given the similarities between very thin black rings and boosted black strings, it is plausible that the nonlinear evolution of the GL instability on thin rings would proceed in a similar manner to that on black strings, thus leading to a violation of WCC in asymptotically flat spaces. This possibility has been contemplated in the past \cite{Elvang:2006dd,Santos:2015iua,Tanabe:2015hda}. Arguably, the resolution of WCC is one of the greatest open problems in classical general relativity, as it directly affects the predictability of the theory.

In this Letter, we report on the end state of black ring instabilities through fully nonlinear, numerical evolution. For very fat rings, the dominant instability is the axisymmetric (``radial'') mode found in Ref. \cite{Figueras:2011he}. Rings with $0.2\lesssim \nu \lesssim 0.6$ are unstable under a new type of nonaxisymmetric instability which deforms the shape of the ring without substantially changing its thickness. In analogy with blackfolds \cite{Emparan:2009at}, we call it an \textit{elastic} mode. In these two regimes, the end point of the instability is the topologically spherical Myers-Perry (MP) black hole. On the other hand, for very thin rings $(\nu\lesssim 0.15)$ the GL instability dominates. Our main focus here will be on thin rings, where our results suggest that the WCC does not hold in the neighborhood of sufficiently thin rings. A more detailed discussion of our results for fatter rings, and a comparison of different angular perturbation modes, will be presented elsewhere \cite{LongPaper}.

%-------------------------------------------------------
% Numerical approach
%-------------------------------------------------------
\PRLsection{Numerical approach}%
 We use the CCZ4 formulation of the five-dimensional Einstein vacuum equations \cite{Alic:2011gg,Weyhausen:2011cg} in Cartesian coordinates $(x,y,z,w)$, with the redefinition of the damping parameter $\kappa_1\to \kappa_1/\alpha$, where $\alpha$ is the lapse \cite{Alic:2013xsa}. We choose $\kappa_1=0.1$ and $\kappa_2 = 0$. We have experimented with other values, but the results do not change. We evolve perturbations of singlyspinning black rings which only break the $U(1)$ symmetry in the $x$-$y$ rotational plane. The remaining $U(1)$ symmetry in the orthogonal $z$-$w$ plane is exploited to dimensionally reduce the CCZ4 equations to $(3+1)$-dimensions using the modified cartoon method \cite{Pretorius:2004jg,Shibata:2010wz}. We do not expect that breaking this orthogonal $U(1)$ symmetry will be relevant in the context of this work.

As initial data, we start with the stationary ring of Ref. \cite{Emparan:2001wn} in the isotropic coordinates introduced in Ref. \cite{Figueras:2011he}. This allows us to transform this solution into Cartesian coordinates. We introduce a small amount of $m=2$ (in the nomenclature of Ref. \cite{Santos:2015iua}) nonaxisymmetric perturbation in the conformal factor $\chi$ via
\begin{equation}
\chi = \chi_0\bigg[1 +A\,\frac{1}{\left(1+Y^2\right)^{\frac{3}{2}}} \frac{x^2-y^2}{\Sigma}\,\bigg]\,,\\
\label{eq:initdat}
\end{equation}
where $\chi_0$ is the unperturbed conformal factor of the stationary black ring, $A$ is the perturbation amplitude, and
\begin{align}
\Sigma=\sqrt{(\tilde R^2+r^2)^2 - 4\,\tilde R^2\,(x^2+y^2)}\,,
\; r = \sqrt{x^2 + y^2 + z^2}\, \nonumber \\
Y = \frac{4(1-\nu)\Sigma}{\nu(r^2+\tilde R^2-\Sigma)}\,, \; \;
\tilde{R} = R \sqrt{\left( 1-\nu \vphantom{\nu^2} \right) / \left( 1+\nu^2 \vphantom{\nu^2} \right)} \, .
\label{eq:pert}
\end{align}
Here $0 < \nu < 1$ and $R > 0$ are the ring's thickness and radius parameters, respectively. In our simulations, we fix $\tilde R=1$ and vary $\nu$. This ensures that the initial coordinate radius of the black ring is roughly 1 for all values of $\nu$, but the mass and hence the instability time scale will vary.

Our initial condition violates the Hamiltonian and momentum constraint equations. However, by using small values of $A$, we can ensure that constraint violations in the initial data are correspondingly small. These small constraint violations are quickly suppressed by the damping terms in the CCZ4 equations. In our simulations, we choose $10^{-6} \leq A \leq 0.002$. The radial dependence of the perturbation \eqref{eq:initdat} is chosen to ensure that it is localized on the horizon and therefore does not change the mass nor the angular momentum of the background spacetime.

In our coordinates, $\Sigma=0$ is a coordinate singularity that corresponds to another asymptotically flat region at the other side of the Einstein-Rosen bridge. We regulate this singularity using the ``turduckening" approach \cite{Brown:2007pg,Brown:2008sb} by manually restricting to $\Sigma \geq \epsilon^2$, for some small $\epsilon$. We choose $\epsilon$ such that the width of the region in which $\Sigma$ is modified is at most $50\%$ of the unperturbed ring's horizon.

To evolve the lapse, we use the CCZ4 $(1+\log)$ slicing \cite{Alic:2011gg} with an advection term, starting from the initial condition $\alpha = \chi$. However, we could not use the standard Gamma-Driver shift condition \cite{Alcubierre:2003} as it quickly freezes the large initial values of $\tilde \Gamma^i$,  even with advection terms.  Instead, we evolve the shift using
\begin{align}
\partial_t \beta^i = F(\tilde \Gamma^i - f(t) \tilde \Gamma^i_{t=0}) - \eta (\beta^i - \beta^i_{t=0}) + \beta^k
\partial_k \beta^i,
\end{align}
where $\tilde \Gamma^i$ is the evolved conformal connection function and
\begin{align}
f(t) = \exp \left[ - \left(\delta_1Y^2 + \delta_2\right)
t^2/M \right]\,,
\end{align}
$Y$ is as defined in Eq. (\ref{eq:pert}), $\delta_1$ and $\delta_2$ are dimensionless parameters, and $M$ is the mass of the unperturbed ring. For our simulations we use $F=2/3$, $\eta = 1$, $\delta_1 = 0.25$, and $\delta_2 = 0.1$. The initial shift is taken to be $\chi$ times the analytic shift.

We evolve the CCZ4 equations numerically on an adaptively refined mesh using the \textsc{GRChombo} code \cite{Clough:2015sqa,chombo-design-doc}. We discretize the equations in space using fourth order finite differences and integrate in time with RK4. We use between 8 to 13 levels of refinement depending on the thickness of the ring. The finest resolution is chosen such that the interior of the horizon is never covered by less than $50$ grid points after gauge adjustment. At the outer boundaries we impose periodic boundary conditions. However, the spatial extent of the domain is made sufficiently large so as to avoid spurious boundary effects throughout the course of the simulation.

To stop the formation of large gradients in $\tilde \gamma_{ij}$ close to the ring singularity, we add a new diffusion term to
the CCZ4 equations, which is restricted to act only inside a region amounting to less than 50\% of the horizon's interior. This is reminiscent of \emph{shock-capturing} techniques in computational fluid dynamics \cite{lapidus}. The additional term does not change the evolution outside the horizon since we have enough grid points across the horizon and the diffusion term only affects features at very small scales. See the Supplemental Material \cite{Supplemental} for more details.

%-------------------------------------------------------
% Results
%-------------------------------------------------------
\PRLsection{Results}%
For rings with $0.3\lesssim \nu \lesssim 0.6$, we find that the evolution is dominated by a new nonaxisymmetric mode which is distinct from the GL mode identified in Ref. \cite{Santos:2015iua}. Note that this range includes both thin and fat rings.
In the nonlinear regime, this new mode deforms the ring without substantially changing its thickness.  We identify it as an elastic mode. In Fig.~\ref*{fig:AH}(a) we display a snapshot of the apparent horizon for a ring with $\nu=0.4$ in the highly nonlinear regime of the evolution.  The deformation caused by the elastic mode can also be seen in Fig.~\ref{fig:AH} (top left). The divergence between the maximum and minimum $S^1$ radii shows that the ring is physically stretching. To measure the influence of the GL mode, we look at the degree of nonuniformity \emph{along} the ring by plotting the maximum and minimum radius of the $S^2$ of the ring as measured by cross-sectional area. The result is shown in Fig.~\ref{fig:AH} (top right). For rings in this range of $\nu$, the minimum $S^2$ radius never decreases substantially, and the growth rate of the elastic mode is larger than the GL mode. The latter is therefore completely irrelevant as far as the nonlinear dynamics is concerned. In fact, the growth rate of the GL mode decreases as the rings become fatter, and for $0.4\lesssim \nu \lesssim 0.6$ the complete gravitational waveforms show that only the elastic mode is relevant. This new instability always ends in a collapse into a topologically spherical MP black hole.
\begin{figure}[b]
\includegraphics[scale=0.85]{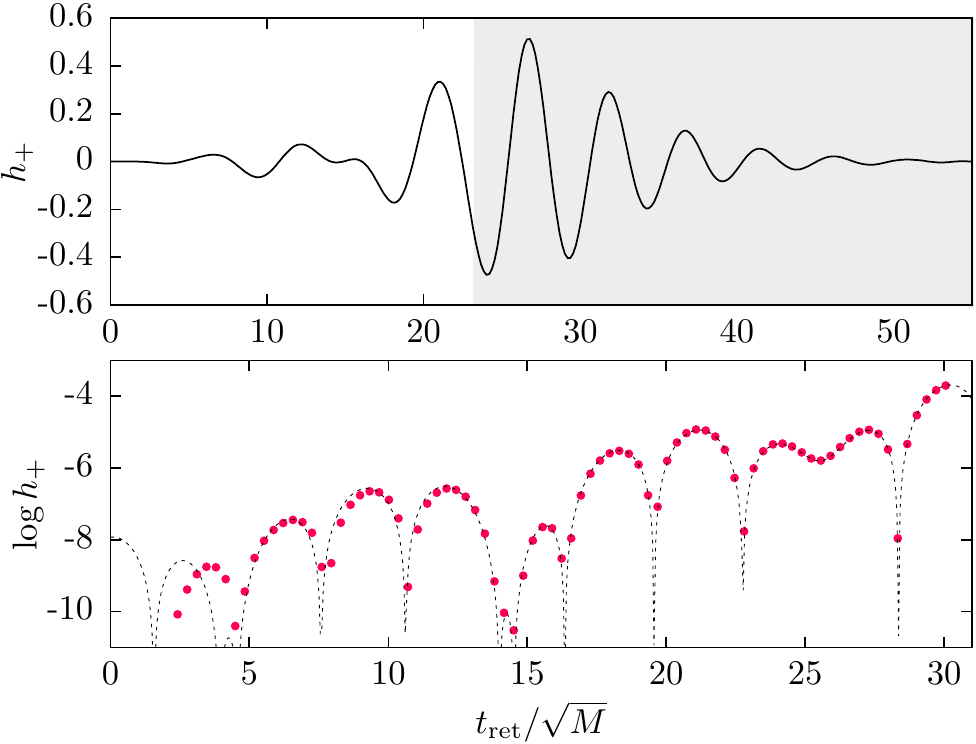}
\includegraphics[scale=0.8]{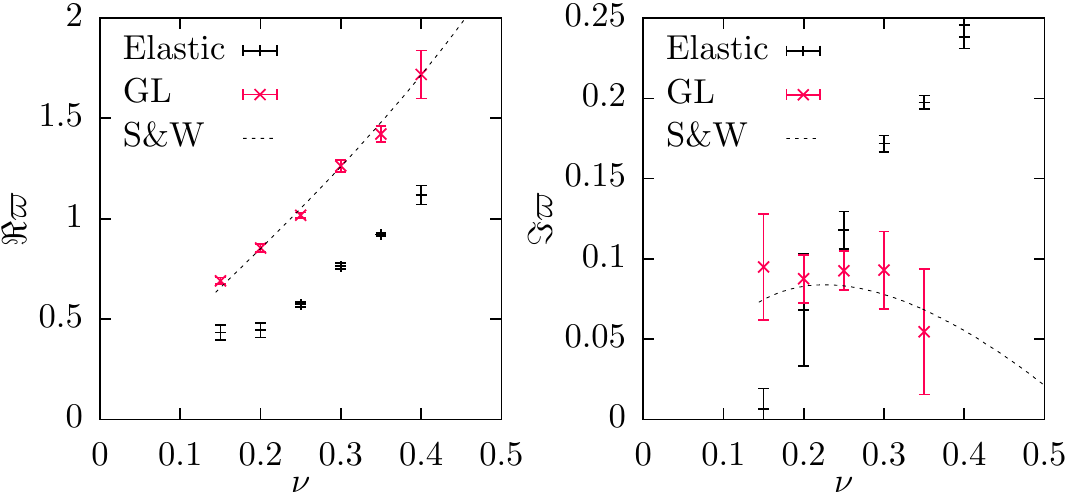}
\caption{Top: complete gravitational waveform for the evolution of the $\nu=0.25$ ring perturbed with an $m=2$
mode with amplitude $A=5\times10^{-4}$. The shaded part corresponds to the portion of the evolution where the outermost
apparent horizon has spherical topology. Middle: fit \eqref{eq:fit} 
of the actual data in the linear regime (red dots) for a perturbation with amplitude $A=10^{-5}$. At the early stages of the evolution there
is contamination from constraint violating modes. Bottom: Real (left) and imaginary (right) parts of the
frequency, $\varpi \equiv \omega/(2\pi\,T)$, of the gravitational waves in the linear regime.
Here, $T$ is the temperature of the unperturbed ring.
The dashed lines correspond to the results of Ref. \cite{Santos:2015iua}. For $\nu=0.4$ we could not reliably extract the growth rate of the GL mode.}
\label{fig:waves}
\end{figure}
\begin{figure}[b]
\includegraphics[scale=0.7]{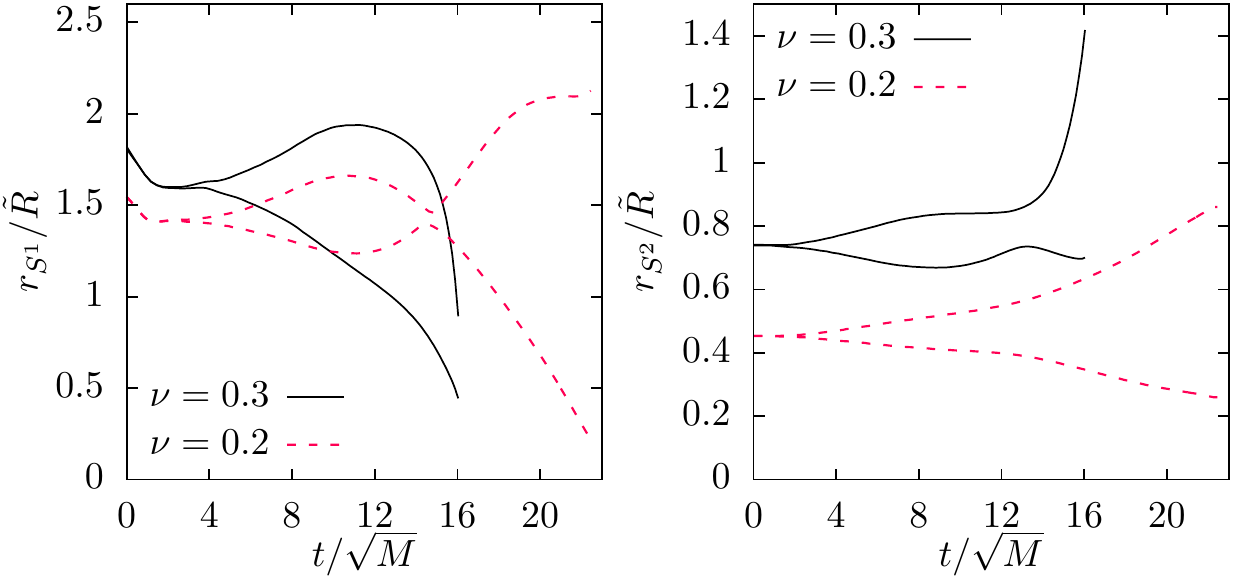}
\begin{overpic}[scale=.11,unit=1mm]{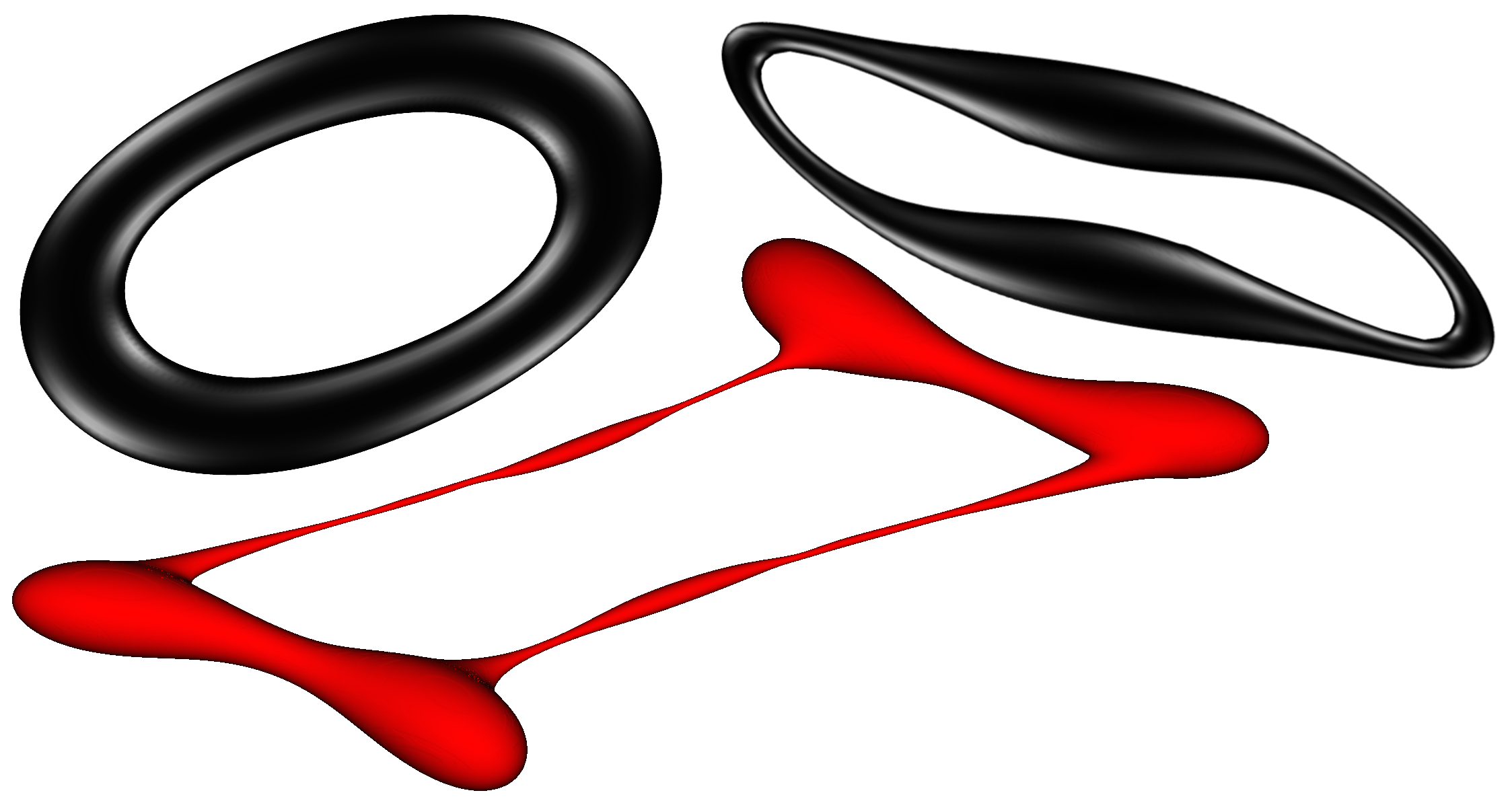}
  \put(4,41){(a)}
  \put(75,41){(b)}
  \put(37,5){(c)}
\end{overpic}
\caption{Top left: maximum and minimum inner radius of the $S^1$ of the ring, as measured by the geodesic distance from the center of the ring to the apparent horizon. For $\nu=0.2$ the maximum and the minimum eventually switch due to the displacement of the bulges. Top right: maximum and minimum areal radius of the $S^2$ of the ring. (a) Apparent horizon of a $\nu=0.4$ ring in the highly dynamical stages of the evolution. (b) Apparent horizon of the $\nu=0.2$ ring just before the collapse into a spherical black hole. (c) $\chi=0.2$ contour for a $\nu=0.15$ ring during the evolution.}
\label{fig:AH}
\end{figure}
For thin rings with $0.2\lesssim\nu\lesssim 0.35$, we observe a competition between different modes.
The waveform in Fig.~\ref{fig:waves} (top) shows that, in the linear regime, there is an apparent mode mixing until nonlinearities become important.  To gain a better understanding of the various modes in the $m=2$ sector of nonaxisymmetric gravitational perturbations, we extract the waveforms by monitoring 
\begin{equation}
h_{+} = \frac{\tilde\gamma_{xx} - \tilde\gamma_{yy}}{2}\,\left(\frac{r}{\tilde R}\right)^{\frac{3}{2}}\,,
\end{equation}
along the $z$ axis. From this, we can identify the frequencies and growth rates of the two modes by fitting the data to
\begin{equation}
A_1\sin(\Re\varpi_1 t+\varphi_1)e^{\Im\varpi_1 t} + A_2\sin(\Re\varpi_2 t+\varphi_2)e^{\Im\varpi_2 t}.
\label{eq:fit}
\end{equation}
We give further details about our fitting procedure and error estimation in the Supplemental Material \cite{Supplemental}. In Fig.~\ref{fig:waves} (middle) we compare the data with the fit \eqref{eq:fit} to show that they are in very good agreement. This confirms that the linear dynamics is governed by the two modes that we have identified. In Fig.~\ref{fig:waves} (bottom) we display the frequencies and growth rates of both the elastic and GL $m=2$ modes. Our results for the GL mode agree very well with Ref. \cite{Santos:2015iua}; however, we were only able to identify the GL mode for thin enough rings $(\nu\lesssim 0.4)$. For thicker rings, the growth rate of the GL mode is much slower than that of the elastic mode, and hence it is much harder to isolate in a fully nonlinear evolution. The fitting procedure also allows us to estimate the amplitude of each mode in our initial data \eqref{eq:initdat}. Both $m=2$ modes have comparable amplitudes initially, and therefore, our simulation is not biased towards the newly identified elastic mode.

Since both modes have similar growth rates for $0.2\lesssim\nu\lesssim 0.35$, it is not surprising that the nonlinear dynamics is governed by a combination of the two.  Figure~\ref{fig:AH} (top left) shows a significant divergence in $S^1$ radii for a $\nu=0.2$ ring on a much larger scale than the ring's thickness. This is indicative of the elastic mode dynamics. On the other hand, in Fig.~\ref{fig:AH} (top right) we observe that GL dynamics causes this ring to also become highly nonuniform. The combined effect of these two modes on the apparent horizon is shown in Fig.~2\ref*{fig:AH}(b). Even though the GL mode does grow significantly here, the end point is still a MP black hole. Presumably, the deformations due to the elastic mode enhance the efficiency of the gravitational wave emission, allowing the ring to quickly shed angular momentum and mass and collapse into a spherical black hole. Therefore, no violation of WCC is observed in this particular case.  

For rings with $\nu\lesssim 0.15$, the GL mode grows fastest and thus dominates the dynamics. In this regime, we consider the nonlinear evolution of a $\nu=0.15$ ring with an $m=2$ perturbation with amplitude $A=5\times 10^{-5}$. It turns out that for such thin rings, the $m=4$ mode grows fast enough that excitations from numerical noise also become important. Therefore, we find that the \textit{generic} nonlinear dynamics is governed by a combination of these two modes. In the highly nonlinear regime the apparent horizon consists of big bulges connected by long and thin necks. One would expect that the thin necks should themselves eventually become GL unstable, giving rise to a second generation of bulges connected by even thinner necks. For such highly deformed dynamical rings, the apparent
horizon is no longer a single-valued function $Y:\,S^1\times S^2\to\mathbb R$, causing our horizon finder to fail (see Supplemental Material \cite{Supplemental}). However, in our gauge the apparent horizon follows certain contours of
the conformal factor, $\chi$. We use these as an indication of the location and shape of the apparent horizon in lieu of the actual surface. In Fig.~2\ref*{fig:AH}(c) we display the $\chi=0.2$ contour for the $\nu=0.15$ ring for an $m=2$ perturbation with $A=5\times 10^{-5}$ at $t/\sqrt{M}=33.87$. This shows clear evidence that a new generation of bulges has formed along the thin necks. We could not continue the evolution due to the limitation in our computational resources, but our results provide enough evidence that this instability should continue in a qualitatively similar manner as in the static black string.  More precisely, the horizon should develop a fractal
structure consisting of big bulges connected by thin necks at different scales. The thinnest necks should reach zero size, and hence a naked singularity should form, in finite asymptotic time. Since there is no fine-tuning involved, this result provides evidence that WCC is violated near thin enough black ring spacetimes. We note that a pure $m=4$ perturbation also gives rise to a similar structure. However, without the stretching effect from $m=2$ the instability's time scale is much longer as the necks are shorter. Significant additional resources will be required to reach the second generation of bulges in this case; however, there is no reason to expect that the end point should be any different. The $m=1$ and higher $m$ modes are also unstable and their study will be presented elsewhere \cite{LongPaper}. 

%-------------------------------------------------------
% Conclusions
%-------------------------------------------------------
\PRLsection{Conclusions}%
We have studied the nonlinear dynamics of thin and fat black rings under nonaxisymmetric perturbations. For rings with $\nu\gtrsim 0.2$ the end point of the instabilities is a MP black hole with a lower angular momentum than the original ring. On the other hand, the GL instability dominates the evolution of thin enough ($\nu\lesssim 0.15$) rings, and the end point should be the pinch-off of the ring. This indicates that WCC is violated around these black ring spacetimes. Note that for these rings the dimensionless angular momentum \cite{Emparan:2006mm} is not particularly large, $j\sim 1.12$. Therefore, our results suggest that violations of WCC can occur for asymptotically flat black holes with $j$ of order 1. Even though we have only considered the $D=5$ case, this conclusion should extend to higher dimensions.  

We have also identified a new, elastic-type of instability in five-dimensional black rings.
This had not been anticipated and was not seen in Refs. \cite{Santos:2015iua,Tanabe:2015hda}.
However, it plays a crucial role in the end point of generic nonaxisymmetric instabilities as it dominates for rings with $\nu\gtrsim 0.2$. It would be very interesting to do a more thorough analysis of the nonaxisymmetric gravitational perturbations of black rings and get a detailed understanding of these elastic instabilities.  

Finally, we introduced a robust and simple new method, based on localized diffusion, to handle singularities in numerical general relativity. While it is used in conjunction with the moving puncture method in the present work, we anticipate that it has a wider range of applications. We will present a more detailed analysis of this in
future work \cite{DiffusionPaper}.

%-------------------------------------------------------
% Acknowledgements
%-------------------------------------------------------
%\PRLsection{Acknowledgements}%
\nocite{Sperhake:2012me}
\nocite{Zilhao:2010sr,Witek:2013koa}
\nocite{Shibata:2010wz}
\nocite{donea,lapidus,vonNeumann}
\nocite{AHFinder}
\nocite{Emparan:2001wn,Emparan:2006mm}
\nocite{petsc-efficient,petsc-user-ref}
\nocite{Meijering99imagereconstruction}
\nocite{Dias:2014eua}
We are very grateful to Garth Wells (Dept. Engineering, U. Cambridge) for suggesting to us the shock capturing technique which has proven so valuable in this work. We would like to thank  J. Briggs, J. Camps, R. Emparan, J. J\"{a}ykk\"{a}, K. Kornet, L. Lehner, F. Pretorius, H. Reall, E. Schnetter, U. Sperhake, T. Wiseman and H. Witek  for numerous stimulating discussions. P.F.  would like to especially thank E. Schnetter and U. Sperhake for early collaboration
in this project. We are very grateful to our collaborators and co-developers of the \textsc{GRChombo} code, K. Clough,
E. Lim and H. Finkel. We would also like to thank J. Santos and B. Way for allowing us to display their data in Fig.~\ref{fig:waves}.  A significant part of this work was undertaken on the COSMOS Shared Memory system at DAMTP, University of Cambridge, operated on behalf of the STFC DiRAC HPC Facility. This equipment is funded by BIS National E-infrastructure capital Grant No. ST/J005673/1 and STFC Grants No. ST/H008586/1, No. ST/K00333X/1. Further portions of this research were conducted with high performance computational resources provided by Louisiana State University \cite{lsu} on its SuperMike-II cluster under allocation NUMREL06. The authors also acknowledge HPC resources from the NSF-XSEDE Grant No. PHY-090003, provided by the Texas Advanced Computing Center (TACC) at The University of Texas at Austin on its Stampede cluster, and by the San Diego Supercomputer Center (SDSC) at UC San Diego on its Comet cluster. P.F. and S.T. were supported by the European Research Council Grant No. ERC-2011-StG 279363-HiDGR. P.F. was
also supported by the Stephen Hawking Advanced Research Fellowship from the Centre
for Theoretical Cosmology, University of Cambridge. P.F. is currently supported by a Royal Society University Research Fellowship and by the European Research Council Grant No. ERC-2014-StG 639022-NewNGR. MK is supported by an STFC studentship. P.F. wants to thank Perimeter Institute and Princeton University for hospitality during various stages of this work.
\bibliography{refs}

%merlin.mbs apsrev4-1.bst 2010-07-25 4.21a (PWD, AO, DPC) hacked
%Control: key (0)
%Control: author (72) initials jnrlst
%Control: editor formatted (1) identically to author
%Control: production of article title (-1) disabled
%Control: page (0) single
%Control: year (1) truncated
%Control: production of eprint (0) enabled
\begin{thebibliography}{42}%
\makeatletter
\providecommand \@ifxundefined [1]{%
 \@ifx{#1\undefined}
}%
\providecommand \@ifnum [1]{%
 \ifnum #1\expandafter \@firstoftwo
 \else \expandafter \@secondoftwo
 \fi
}%
\providecommand \@ifx [1]{%
 \ifx #1\expandafter \@firstoftwo
 \else \expandafter \@secondoftwo
 \fi
}%
\providecommand \natexlab [1]{#1}%
\providecommand \enquote  [1]{``#1''}%
\providecommand \bibnamefont  [1]{#1}%
\providecommand \bibfnamefont [1]{#1}%
\providecommand \citenamefont [1]{#1}%
\providecommand \href@noop [0]{\@secondoftwo}%
\providecommand \href [0]{\begingroup \@sanitize@url \@href}%
\providecommand \@href[1]{\@@startlink{#1}\@@href}%
\providecommand \@@href[1]{\endgroup#1\@@endlink}%
\providecommand \@sanitize@url [0]{\catcode `\\12\catcode `\$12\catcode
  `\&12\catcode `\#12\catcode `\^12\catcode `\_12\catcode `\%12\relax}%
\providecommand \@@startlink[1]{}%
\providecommand \@@endlink[0]{}%
\providecommand \url  [0]{\begingroup\@sanitize@url \@url }%
\providecommand \@url [1]{\endgroup\@href {#1}{\urlprefix }}%
\providecommand \urlprefix  [0]{URL }%
\providecommand \Eprint [0]{\href }%
\providecommand \doibase [0]{http://dx.doi.org/}%
\providecommand \selectlanguage [0]{\@gobble}%
\providecommand \bibinfo  [0]{\@secondoftwo}%
\providecommand \bibfield  [0]{\@secondoftwo}%
\providecommand \translation [1]{[#1]}%
\providecommand \BibitemOpen [0]{}%
\providecommand \bibitemStop [0]{}%
\providecommand \bibitemNoStop [0]{.\EOS\space}%
\providecommand \EOS [0]{\spacefactor3000\relax}%
\providecommand \BibitemShut  [1]{\csname bibitem#1\endcsname}%
\let\auto@bib@innerbib\@empty
%</preamble>
\bibitem [{\citenamefont {Kerr}(1963)}]{Kerr:1963ud}%
  \BibitemOpen
  \bibfield  {author} {\bibinfo {author} {\bibfnamefont {R.~P.}\ \bibnamefont
  {Kerr}},\ }\href {\doibase 10.1103/PhysRevLett.11.237} {\bibfield  {journal}
  {\bibinfo  {journal} {Phys. Rev. Lett.}\ }\textbf {\bibinfo {volume} {11}},\
  \bibinfo {pages} {237} (\bibinfo {year} {1963})}\BibitemShut {NoStop}%
%%CITATION = PRLTA,11,237;%%
\bibitem [{\citenamefont {Whiting}(1989)}]{Whiting:1988vc}%
  \BibitemOpen
  \bibfield  {author} {\bibinfo {author} {\bibfnamefont {B.~F.}\ \bibnamefont
  {Whiting}},\ }\href {\doibase 10.1063/1.528308} {\bibfield  {journal}
  {\bibinfo  {journal} {J. Math. Phys. (N.Y.)}\ }\textbf {\bibinfo {volume} {30}},\
  \bibinfo {pages} {1301} (\bibinfo {year} {1989})}\BibitemShut {NoStop}%
%%CITATION = JMAPA,30,1301;%%
\bibitem [{\citenamefont {Dafermos}\ and\ \citenamefont
  {Rodnianski}()}]{Dafermos:2010hb}%
  \BibitemOpen
  \bibfield  {author} {\bibinfo {author} {\bibfnamefont {M.}~\bibnamefont
  {Dafermos}}\ and\ \bibinfo {author} {\bibfnamefont {I.}~\bibnamefont
  {Rodnianski}},\ }\href@noop {} {\ }\Eprint {http://arxiv.org/abs/1010.5132}
  {arXiv:1010.5132} \BibitemShut {NoStop}%
%%CITATION = ARXIV:1010.5132;%%
\bibitem [{\citenamefont {Dafermos}\ \emph
  {et~al.}(2014{\natexlab{a}})\citenamefont {Dafermos}, \citenamefont
  {Rodnianski},\ and\ \citenamefont {Shlapentokh-Rothman}}]{Dafermos:2014cua}%
  \BibitemOpen
  \bibfield  {author} {\bibinfo {author} {\bibfnamefont {M.}~\bibnamefont
  {Dafermos}}, \bibinfo {author} {\bibfnamefont {I.}~\bibnamefont
  {Rodnianski}}, \ and\ \bibinfo {author} {\bibfnamefont {Y.}~\bibnamefont
  {Shlapentokh-Rothman}},\ }\href@noop {} {\  (\bibinfo {year}
  {2014}{\natexlab{a}})},\ \Eprint {http://arxiv.org/abs/1402.7034}
  {arXiv:1402.7034} \BibitemShut {NoStop}%
%%CITATION = ARXIV:1402.7034;%%
\bibitem [{\citenamefont {Dafermos}\ \emph
  {et~al.}(2014{\natexlab{b}})\citenamefont {Dafermos}, \citenamefont
  {Rodnianski},\ and\ \citenamefont {Shlapentokh-Rothman}}]{Dafermos:2014jwa}%
  \BibitemOpen
  \bibfield  {author} {\bibinfo {author} {\bibfnamefont {M.}~\bibnamefont
  {Dafermos}}, \bibinfo {author} {\bibfnamefont {I.}~\bibnamefont
  {Rodnianski}}, \ and\ \bibinfo {author} {\bibfnamefont {Y.}~\bibnamefont
  {Shlapentokh-Rothman}},\ }\href@noop {} {\  (\bibinfo {year}
  {2014}{\natexlab{b}})},\ \Eprint {http://arxiv.org/abs/1412.8379}
  {arXiv:1412.8379} \BibitemShut {NoStop}%
%%CITATION = ARXIV:1412.8379;%%
\bibitem [{\citenamefont {Gregory}\ and\ \citenamefont
  {Laflamme}(1993)}]{Gregory:1993vy}%
  \BibitemOpen
  \bibfield  {author} {\bibinfo {author} {\bibfnamefont {R.}~\bibnamefont
  {Gregory}}\ and\ \bibinfo {author} {\bibfnamefont {R.}~\bibnamefont
  {Laflamme}},\ }\href {\doibase 10.1103/PhysRevLett.70.2837} {\bibfield
  {journal} {\bibinfo  {journal} {Phys. Rev. Lett.}\ }\textbf {\bibinfo
  {volume} {70}},\ \bibinfo {pages} {2837} (\bibinfo {year}
  {1993})}\BibitemShut {NoStop}%
%%CITATION = HEP-TH/9301052;%%
\bibitem [{\citenamefont {Hovdebo}\ and\ \citenamefont
  {Myers}(2006)}]{Hovdebo:2006jy}%
  \BibitemOpen
  \bibfield  {author} {\bibinfo {author} {\bibfnamefont {J.~L.}\ \bibnamefont
  {Hovdebo}}\ and\ \bibinfo {author} {\bibfnamefont {R.~C.}\ \bibnamefont
  {Myers}},\ }\href {\doibase 10.1103/PhysRevD.73.084013} {\bibfield  {journal} {\bibinfo  {journal} {Phys.
  Rev.}\ }\textbf {\bibinfo {volume} {D 73}},\ \bibinfo {pages} {084013}
  (\bibinfo {year} {2006})}\BibitemShut {NoStop}%
%%CITATION = HEP-TH/0601079;%%
\bibitem [{\citenamefont {Lehner}\ and\ \citenamefont
  {Pretorius}(2010)}]{Lehner:2010pn}%
  \BibitemOpen
  \bibfield  {author} {\bibinfo {author} {\bibfnamefont {L.}~\bibnamefont
  {Lehner}}\ and\ \bibinfo {author} {\bibfnamefont {F.}~\bibnamefont
  {Pretorius}},\ }\href {\doibase 10.1103/PhysRevLett.105.101102} {\bibfield
  {journal} {\bibinfo  {journal} {Phys. Rev. Lett.}\ }\textbf {\bibinfo
  {volume} {105}},\ \bibinfo {pages} {101102} (\bibinfo {year}
  {2010})}\BibitemShut {NoStop}%
%%CITATION = ARXIV:1006.5960;%%
\bibitem [{\citenamefont {Penrose}(1969)}]{Penrose:1969pc}%
  \BibitemOpen
  \bibfield  {author} {\bibinfo {author} {\bibfnamefont {R.}~\bibnamefont
  {Penrose}},\ }\href@noop {} {\bibfield  {journal} {\bibinfo  {journal} {Riv.
  Nuovo Cim.}\ }\textbf {\bibinfo {volume} {1}},\ \bibinfo {pages} {252}
  (\bibinfo {year} {1969})},\ \bibinfo {note} {[\href{\doibase 10.1023/A:1016578408204}{Gen. Relativ.
  Gravit. \textbf{34}, 1141 (2002)}]}\BibitemShut {NoStop}%
%%CITATION = RNCIB,1,252;%%
\bibitem [{\citenamefont {Christodoulou}(1999)}]{Christodoulou:1999ve}%
  \BibitemOpen
  \bibfield  {author} {\bibinfo {author} {\bibfnamefont {D.}~\bibnamefont
  {Christodoulou}},\ }\href {\doibase 10.1088/0264-9381/16/12A/302} {\bibfield
   {journal} {\bibinfo  {journal} {Classical and Quantum Gravity}\ }\textbf
  {\bibinfo {volume} {16}},\ \bibinfo {pages} {A23} (\bibinfo {year}
  {1999})}\BibitemShut {NoStop}%
\bibitem [{\citenamefont {Emparan}\ and\ \citenamefont
  {Reall}(2002)}]{Emparan:2001wn}%
  \BibitemOpen
  \bibfield  {author} {\bibinfo {author} {\bibfnamefont {R.}~\bibnamefont
  {Emparan}}\ and\ \bibinfo {author} {\bibfnamefont {H.~S.}\ \bibnamefont
  {Reall}},\ }\href {\doibase 10.1103/PhysRevLett.88.101101} {\bibfield
  {journal} {\bibinfo  {journal} {Phys. Rev. Lett.}\ }\textbf {\bibinfo
  {volume} {88}},\ \bibinfo {pages} {101101} (\bibinfo {year}
  {2002})}\BibitemShut {NoStop}%
%%CITATION = HEP-TH/0110260;%%
\bibitem [{\citenamefont {Emparan}\ and\ \citenamefont
  {Reall}(2006)}]{Emparan:2006mm}%
  \BibitemOpen
  \bibfield  {author} {\bibinfo {author} {\bibfnamefont {R.}~\bibnamefont
  {Emparan}}\ and\ \bibinfo {author} {\bibfnamefont {H.~S.}\ \bibnamefont
  {Reall}},\ }\href {\doibase 10.1088/0264-9381/23/20/R01} {\bibfield
  {journal} {\bibinfo  {journal} {Classical and Quantum Gravity}\ }\textbf
  {\bibinfo {volume} {23}},\ \bibinfo {pages} {R169} (\bibinfo {year}
  {2006})}\BibitemShut {NoStop}%
%%CITATION = HEP-TH/0608012;%%
\bibitem [{\citenamefont {Elvang}\ \emph {et~al.}(2006)\citenamefont {Elvang},
  \citenamefont {Emparan},\ and\ \citenamefont {Virmani}}]{Elvang:2006dd}%
  \BibitemOpen
  \bibfield  {author} {\bibinfo {author} {\bibfnamefont {H.}~\bibnamefont
  {Elvang}}, \bibinfo {author} {\bibfnamefont {R.}~\bibnamefont {Emparan}}, \
  and\ \bibinfo {author} {\bibfnamefont {A.}~\bibnamefont {Virmani}},\ }\href
  {\doibase 10.1088/1126-6708/2006/12/074} {\bibfield  {journal} {\bibinfo
  {journal} {J. High Energy Phys.}\ }\textbf {\bibinfo {volume} {12}},\
  \bibinfo {pages} {074} (\bibinfo {year} {2006})}\BibitemShut {NoStop}%
%%CITATION = HEP-TH/0608076;%%
\bibitem [{\citenamefont {Figueras}\ \emph {et~al.}(2011)\citenamefont
  {Figueras}, \citenamefont {Murata},\ and\ \citenamefont
  {Reall}}]{Figueras:2011he}%
  \BibitemOpen
  \bibfield  {author} {\bibinfo {author} {\bibfnamefont {P.}~\bibnamefont
  {Figueras}}, \bibinfo {author} {\bibfnamefont {K.}~\bibnamefont {Murata}}, \
  and\ \bibinfo {author} {\bibfnamefont {H.~S.}\ \bibnamefont {Reall}},\ }\href
  {\doibase 10.1088/0264-9381/28/22/225030} {\bibfield  {journal} {\bibinfo
  {journal} {Classical and Quantum Gravity}\ }\textbf {\bibinfo {volume}
  {28}},\ \bibinfo {pages} {225030} (\bibinfo {year} {2011})}\BibitemShut
  {NoStop}%
%%CITATION = ARXIV:1107.5785;%%
\bibitem [{\citenamefont {Santos}\ and\ \citenamefont
  {Way}(2015)}]{Santos:2015iua}%
  \BibitemOpen
  \bibfield  {author} {\bibinfo {author} {\bibfnamefont {J.~E.}\ \bibnamefont
  {Santos}}\ and\ \bibinfo {author} {\bibfnamefont {B.}~\bibnamefont {Way}},\
  }\href {\doibase 10.1103/PhysRevLett.114.221101} {\bibfield  {journal}
  {\bibinfo  {journal} {Phys. Rev. Lett.}\ }\textbf {\bibinfo {volume} {114}},\
  \bibinfo {pages} {221101} (\bibinfo {year} {2015})}\BibitemShut {NoStop}%
%%CITATION = ARXIV:1503.00721;%%
\bibitem [{\citenamefont {Tanabe}()}]{Tanabe:2015hda}%
  \BibitemOpen
  \bibfield  {author} {\bibinfo {author} {\bibfnamefont {K.}~\bibnamefont
  {Tanabe}},\ }\href@noop {} {\ }\Eprint {http://arxiv.org/abs/1510.02200}
  {arXiv:1510.02200} \BibitemShut {NoStop}%
%%CITATION = ARXIV:1510.02200;%%
\bibitem [{\citenamefont {Emparan}\ \emph {et~al.}(2010)\citenamefont
  {Emparan}, \citenamefont {Harmark}, \citenamefont {Niarchos},\ and\
  \citenamefont {Obers}}]{Emparan:2009at}%
  \BibitemOpen
  \bibfield  {author} {\bibinfo {author} {\bibfnamefont {R.}~\bibnamefont
  {Emparan}}, \bibinfo {author} {\bibfnamefont {T.}~\bibnamefont {Harmark}},
  \bibinfo {author} {\bibfnamefont {V.}~\bibnamefont {Niarchos}}, \ and\
  \bibinfo {author} {\bibfnamefont {N.~A.}\ \bibnamefont {Obers}},\ }\href
  {\doibase 10.1007/JHEP03(2010)063} {\bibfield  {journal} {\bibinfo  {journal}
  {J. High Energy Phys.}\ }\textbf {\bibinfo {volume} {03}},\ \bibinfo {pages}
  {063} (\bibinfo {year} {2010})}\BibitemShut {NoStop}%
%%CITATION = ARXIV:0910.1601;%%
\bibitem [{\citenamefont {Figueras}\ \emph
  {et~al.}({\natexlab{a}})\citenamefont {Figueras}, \citenamefont {Kunesch},\
  and\ \citenamefont {Tunyasuvunakool}}]{LongPaper}%
  \BibitemOpen
  \bibfield  {author} {\bibinfo {author} {\bibfnamefont {P.}~\bibnamefont
  {Figueras}}, \bibinfo {author} {\bibfnamefont {M.}~\bibnamefont {Kunesch}}, \
  and\ \bibinfo {author} {\bibfnamefont {S.}~\bibnamefont {Tunyasuvunakool}} \
  }\href@noop {} {\bibfield  {journal} 
  (to be published)}\BibitemShut {NoStop}%
\bibitem [{\citenamefont {Alic}\ \emph {et~al.}(2012)\citenamefont {Alic},
  \citenamefont {Bona-Casas}, \citenamefont {Bona}, \citenamefont {Rezzolla},\
  and\ \citenamefont {Palenzuela}}]{Alic:2011gg}%
  \BibitemOpen
  \bibfield  {author} {\bibinfo {author} {\bibfnamefont {D.}~\bibnamefont
  {Alic}}, \bibinfo {author} {\bibfnamefont {C.}~\bibnamefont {Bona-Casas}},
  \bibinfo {author} {\bibfnamefont {C.}~\bibnamefont {Bona}}, \bibinfo {author}
  {\bibfnamefont {L.}~\bibnamefont {Rezzolla}}, \ and\ \bibinfo {author}
  {\bibfnamefont {C.}~\bibnamefont {Palenzuela}},\ }\href {\doibase
  10.1103/PhysRevD.85.064040} {\bibfield  {journal} {\bibinfo  {journal} {Phys.
  Rev.}\ }\textbf {\bibinfo {volume} {D 85}},\ \bibinfo {pages} {064040}
  (\bibinfo {year} {2012})}\BibitemShut {NoStop}%
%%CITATION = ARXIV:1106.2254;%%
\bibitem [{\citenamefont {Weyhausen}\ \emph {et~al.}(2012)\citenamefont
  {Weyhausen}, \citenamefont {Bernuzzi},\ and\ \citenamefont
  {Hilditch}}]{Weyhausen:2011cg}%
  \BibitemOpen
  \bibfield  {author} {\bibinfo {author} {\bibfnamefont {A.}~\bibnamefont
  {Weyhausen}}, \bibinfo {author} {\bibfnamefont {S.}~\bibnamefont {Bernuzzi}},
  \ and\ \bibinfo {author} {\bibfnamefont {D.}~\bibnamefont {Hilditch}},\
  }\href {\doibase 10.1103/PhysRevD.85.024038} {\bibfield  {journal} {\bibinfo
  {journal} {Phys. Rev.}\ }\textbf {\bibinfo {volume} {D 85}},\ \bibinfo
  {pages} {024038} (\bibinfo {year} {2012})}\BibitemShut {NoStop}%
%%CITATION = ARXIV:1107.5539;%%
\bibitem [{\citenamefont {Alic}\ \emph {et~al.}(2013)\citenamefont {Alic},
  \citenamefont {Kastaun},\ and\ \citenamefont {Rezzolla}}]{Alic:2013xsa}%
  \BibitemOpen
  \bibfield  {author} {\bibinfo {author} {\bibfnamefont {D.}~\bibnamefont
  {Alic}}, \bibinfo {author} {\bibfnamefont {W.}~\bibnamefont {Kastaun}}, \
  and\ \bibinfo {author} {\bibfnamefont {L.}~\bibnamefont {Rezzolla}},\ }\href
  {\doibase 10.1103/PhysRevD.88.064049} {\bibfield  {journal} {\bibinfo
  {journal} {Phys. Rev.}\ }\textbf {\bibinfo {volume} {D 88}},\ \bibinfo
  {pages} {064049} (\bibinfo {year} {2013})}\BibitemShut {NoStop}%
%%CITATION = ARXIV:1307.7391;%%
\bibitem [{\citenamefont {Pretorius}(2005)}]{Pretorius:2004jg}%
  \BibitemOpen
  \bibfield  {author} {\bibinfo {author} {\bibfnamefont {F.}~\bibnamefont
  {Pretorius}},\ }\href {\doibase 10.1088/0264-9381/22/2/014} {\bibfield
  {journal} {\bibinfo  {journal} {Classical and Quantum Gravity}\ }\textbf
  {\bibinfo {volume} {22}},\ \bibinfo {pages} {425} (\bibinfo {year}
  {2005})}\BibitemShut {NoStop}%
%%CITATION = GR-QC/0407110;%%
\bibitem [{\citenamefont {Shibata}\ and\ \citenamefont
  {Yoshino}(2010)}]{Shibata:2010wz}%
  \BibitemOpen
  \bibfield  {author} {\bibinfo {author} {\bibfnamefont {M.}~\bibnamefont
  {Shibata}}\ and\ \bibinfo {author} {\bibfnamefont {H.}~\bibnamefont
  {Yoshino}},\ }\href {\doibase 10.1103/PhysRevD.81.104035} {\bibfield
  {journal} {\bibinfo  {journal} {Phys. Rev.}\ }\textbf {\bibinfo {volume} {D
  81}},\ \bibinfo {pages} {104035} (\bibinfo {year} {2010})}\BibitemShut
  {NoStop}%
%%CITATION = ARXIV:1004.4970;%%
\bibitem [{\citenamefont {Brown}\ \emph {et~al.}(2007)\citenamefont {Brown},
  \citenamefont {Sarbach}, \citenamefont {Schnetter}, \citenamefont {Tiglio},
  \citenamefont {Diener}, \citenamefont {Hawke},\ and\ \citenamefont
  {Pollney}}]{Brown:2007pg}%
  \BibitemOpen
  \bibfield  {author} {\bibinfo {author} {\bibfnamefont {D.}~\bibnamefont
  {Brown}}, \bibinfo {author} {\bibfnamefont {O.}~\bibnamefont {Sarbach}},
  \bibinfo {author} {\bibfnamefont {E.}~\bibnamefont {Schnetter}}, \bibinfo
  {author} {\bibfnamefont {M.}~\bibnamefont {Tiglio}}, \bibinfo {author}
  {\bibfnamefont {P.}~\bibnamefont {Diener}}, \bibinfo {author} {\bibfnamefont
  {I.}~\bibnamefont {Hawke}}, \ and\ \bibinfo {author} {\bibfnamefont
  {D.}~\bibnamefont {Pollney}},\ }\href {\doibase 10.1103/PhysRevD.76.081503}
  {\bibfield  {journal} {\bibinfo  {journal} {Phys. Rev.}\ }\textbf {\bibinfo
  {volume} {D 76}},\ \bibinfo {pages} {081503} (\bibinfo {year}
  {2007})}\BibitemShut {NoStop}%
%%CITATION = ARXIV:0707.3101;%%
\bibitem [{\citenamefont {Brown}\ \emph {et~al.}(2009)\citenamefont {Brown},
  \citenamefont {Diener}, \citenamefont {Sarbach}, \citenamefont {Schnetter},\
  and\ \citenamefont {Tiglio}}]{Brown:2008sb}%
  \BibitemOpen
  \bibfield  {author} {\bibinfo {author} {\bibfnamefont {D.}~\bibnamefont
  {Brown}}, \bibinfo {author} {\bibfnamefont {P.}~\bibnamefont {Diener}},
  \bibinfo {author} {\bibfnamefont {O.}~\bibnamefont {Sarbach}}, \bibinfo
  {author} {\bibfnamefont {E.}~\bibnamefont {Schnetter}}, \ and\ \bibinfo
  {author} {\bibfnamefont {M.}~\bibnamefont {Tiglio}},\ }\href {\doibase
  10.1103/PhysRevD.79.044023} {\bibfield  {journal} {\bibinfo  {journal} {Phys.
  Rev.}\ }\textbf {\bibinfo {volume} {D 79}},\ \bibinfo {pages} {044023}
  (\bibinfo {year} {2009})}\BibitemShut {NoStop}%
%%CITATION = ARXIV:0809.3533;%%
\bibitem [{\citenamefont {Alcubierre}\ \emph {et~al.}(2003)\citenamefont
  {Alcubierre}, \citenamefont {Br\"{u}gmann}, \citenamefont {Diener},
  \citenamefont {Koppitz}, \citenamefont {Pollney}, \citenamefont {Seidel},\
  and\ \citenamefont {Takahashi}}]{Alcubierre:2003}%
  \BibitemOpen
  \bibfield  {author} {\bibinfo {author} {\bibfnamefont {M.}~\bibnamefont
  {Alcubierre}}, \bibinfo {author} {\bibfnamefont {B.}~\bibnamefont
  {Br\"{u}gmann}}, \bibinfo {author} {\bibfnamefont {P.}~\bibnamefont
  {Diener}}, \bibinfo {author} {\bibfnamefont {M.}~\bibnamefont {Koppitz}},
  \bibinfo {author} {\bibfnamefont {D.}~\bibnamefont {Pollney}}, \bibinfo
  {author} {\bibfnamefont {E.}~\bibnamefont {Seidel}}, \ and\ \bibinfo {author}
  {\bibfnamefont {R.}~\bibnamefont {Takahashi}},\ }\href {\doibase
  10.1103/PhysRevD.67.084023} {\bibfield  {journal} {\bibinfo  {journal} {Phys.
  Rev.}\ }\textbf {\bibinfo {volume} {D 67}},\ \bibinfo {pages} {084023}
  (\bibinfo {year} {2003})}\BibitemShut {NoStop}%
%%CITATION = GR-QC/0206072;%%
\bibitem [{\citenamefont {Clough}\ \emph {et~al.}(2015)\citenamefont {Clough},
  \citenamefont {Figueras}, \citenamefont {Finkel}, \citenamefont {Kunesch},
  \citenamefont {Lim},\ and\ \citenamefont {Tunyasuvunakool}}]{Clough:2015sqa}%
  \BibitemOpen
  \bibfield  {author} {\bibinfo {author} {\bibfnamefont {K.}~\bibnamefont
  {Clough}}, \bibinfo {author} {\bibfnamefont {P.}~\bibnamefont {Figueras}},
  \bibinfo {author} {\bibfnamefont {H.}~\bibnamefont {Finkel}}, \bibinfo
  {author} {\bibfnamefont {M.}~\bibnamefont {Kunesch}}, \bibinfo {author}
  {\bibfnamefont {E.~A.}\ \bibnamefont {Lim}}, \ and\ \bibinfo {author}
  {\bibfnamefont {S.}~\bibnamefont {Tunyasuvunakool}},\ }\href {\doibase
  10.1088/0264-9381/32/24/245011} {\bibfield  {journal} {\bibinfo  {journal}
  {Classical and Quantum Gravity}\ }\textbf {\bibinfo {volume} {32}},\ \bibinfo
  {pages} {245011} (\bibinfo {year} {2015})}\BibitemShut {NoStop}%
%%CITATION = ARXIV:1503.03436;%%
\bibitem [{\citenamefont {Adams}\ \emph {et~al.}(2015)\citenamefont {Adams},
  \citenamefont {Colella}, \citenamefont {Graves}, \citenamefont {Johnson},
  \citenamefont {Keen}, \citenamefont {Ligocki}, \citenamefont {Martin},
  \citenamefont {McCorquodale}, \citenamefont {Modiano}, \citenamefont
  {Schwartz}, \citenamefont {Sternberg},\ and\ \citenamefont
  {Van~Straalen}}]{chombo-design-doc}%
  \BibitemOpen
  \bibfield  {author} {\bibinfo {author} {\bibfnamefont {M.}~\bibnamefont
  {Adams}}, \bibinfo {author} {\bibfnamefont {P.}~\bibnamefont {Colella}},
  \bibinfo {author} {\bibfnamefont {D.}~\bibnamefont {Graves}}, \bibinfo
  {author} {\bibfnamefont {J.}~\bibnamefont {Johnson}}, \bibinfo {author}
  {\bibfnamefont {N.}~\bibnamefont {Keen}}, \bibinfo {author} {\bibfnamefont
  {T.}~\bibnamefont {Ligocki}}, \bibinfo {author} {\bibfnamefont
  {D.}~\bibnamefont {Martin}}, \bibinfo {author} {\bibfnamefont
  {P.}~\bibnamefont {McCorquodale}}, \bibinfo {author} {\bibfnamefont
  {D.}~\bibnamefont {Modiano}}, \bibinfo {author} {\bibfnamefont
  {P.}~\bibnamefont {Schwartz}}, \bibinfo {author} {\bibfnamefont
  {T.}~\bibnamefont {Sternberg}}, \ and\ \bibinfo {author} {\bibfnamefont
  {B.}~\bibnamefont {Van~Straalen}},\ }\href
  {http://crd.lbl.gov/assets/pubs_presos/chomboDesign.pdf} {
  \bibinfo{institution} {Lawrence Berkeley National Laboratory}
  \bibinfo {type} {Technical Report}\ No. \bibinfo {number} {LBNL-6616E},\ \bibinfo {year}
  {2015}} (unpublished)\BibitemShut {NoStop}%
\bibitem [{\citenamefont {Lapidus}(1967)}]{lapidus}%
  \BibitemOpen
  \bibfield  {author} {\bibinfo {author} {\bibfnamefont {A.}~\bibnamefont
  {Lapidus}},\ }\href {\doibase 10.1016/0021-9991(67)90032-0} {\bibfield
  {journal} {\bibinfo  {journal} {J. Comput. Phys.}\ }\textbf
  {\bibinfo {volume} {2}},\ \bibinfo {pages} {154 } (\bibinfo {year}
  {1967})}\BibitemShut {NoStop}%
\bibitem [{Sup()}]{Supplemental}%
  \BibitemOpen
  \href@noop {} {}\bibinfo {note} {See Supplemental Material for numerical checks and further technical details,
  which includes Refs. [32-42].}\BibitemShut {Stop}%
\bibitem [{lsu()}]{lsu}%
  \BibitemOpen
  \href {http://www.hpc.lsu.edu} {
  \bibinfo {note} {{http://www.hpc.lsu.edu}}}\BibitemShut {Stop}%
\bibitem [{\citenamefont {Sperhake}\ \emph {et~al.}(2013)\citenamefont
  {Sperhake}, \citenamefont {Berti}, \citenamefont {Cardoso},\ and\
  \citenamefont {Pretorius}}]{Sperhake:2012me}%
  \BibitemOpen
  \bibfield  {author} {\bibinfo {author} {\bibfnamefont {U.}~\bibnamefont
  {Sperhake}}, \bibinfo {author} {\bibfnamefont {E.}~\bibnamefont {Berti}},
  \bibinfo {author} {\bibfnamefont {V.}~\bibnamefont {Cardoso}}, \ and\
  \bibinfo {author} {\bibfnamefont {F.}~\bibnamefont {Pretorius}},\ }\href
  {\doibase 10.1103/PhysRevLett.111.041101} {\bibfield  {journal} {\bibinfo
  {journal} {Phys. Rev. Lett.}\ }\textbf {\bibinfo {volume} {111}},\ \bibinfo
  {pages} {041101} (\bibinfo {year} {2013})}\BibitemShut {NoStop}%
\bibitem [{\citenamefont {Zilhao}\ \emph {et~al.}(2010)\citenamefont {Zilhao},
  \citenamefont {Witek}, \citenamefont {Sperhake}, \citenamefont {Cardoso},
  \citenamefont {Gualtieri}, \citenamefont {Herdeiro},\ and\ \citenamefont
  {Nerozzi}}]{Zilhao:2010sr}%
  \BibitemOpen
  \bibfield  {author} {\bibinfo {author} {\bibfnamefont {M.}~\bibnamefont
  {Zilhao}}, \bibinfo {author} {\bibfnamefont {H.}~\bibnamefont {Witek}},
  \bibinfo {author} {\bibfnamefont {U.}~\bibnamefont {Sperhake}}, \bibinfo
  {author} {\bibfnamefont {V.}~\bibnamefont {Cardoso}}, \bibinfo {author}
  {\bibfnamefont {L.}~\bibnamefont {Gualtieri}}, \bibinfo {author}
  {\bibfnamefont {C.}~\bibnamefont {Herdeiro}}, \ and\ \bibinfo {author}
  {\bibfnamefont {A.}~\bibnamefont {Nerozzi}},\ }\href {\doibase
  10.1103/PhysRevD.81.084052} {\bibfield  {journal} {\bibinfo  {journal} {Phys.
  Rev.}\ }\textbf {\bibinfo {volume} {D 81}},\ \bibinfo {pages} {084052}
  (\bibinfo {year} {2010})}\BibitemShut {NoStop}%
%%CITATION = ARXIV:1001.2302;%%
\bibitem [{\citenamefont {Witek}(2012)}]{Witek:2013koa}%
  \BibitemOpen
  \bibfield  {author} {\bibinfo {author} {\bibfnamefont {H.}~\bibnamefont
  {Witek}},\ }\href@noop {} {Ph.D. thesis},\ \bibinfo  {school} {IST/CENTRA
  Lisbon},\ \bibinfo {year} {2012};\ \Eprint{http://arxiv.org/abs/1307.1145}
  {arXiv:1307.1145}\BibitemShut {NoStop}%
\bibitem [{\citenamefont {Donea}\ and\ \citenamefont {Huerta}(2003)}]{donea}%
  \BibitemOpen
  \bibfield  {author} {\bibinfo {author} {\bibfnamefont {J.}~\bibnamefont
  {Donea}}\ and\ \bibinfo {author} {\bibfnamefont {A.}~\bibnamefont {Huerta}},\
  }\href@noop {} {\emph {\bibinfo {title} {Finite Element Methods for Flow
  Problems}}}\ (\bibinfo  {publisher} {John Wiley \& Sons},\ \bibinfo {year}
  {2003})\BibitemShut {NoStop}%
\bibitem [{\citenamefont {{Von Neumann}}\ and\ \citenamefont
  {{Richtmyer}}(1950)}]{vonNeumann}%
  \BibitemOpen
  \bibfield  {author} {\bibinfo {author} {\bibfnamefont {J.}~\bibnamefont {{Von
  Neumann}}}\ and\ \bibinfo {author} {\bibfnamefont {R.~D.}\ \bibnamefont
  {{Richtmyer}}},\ }\href {\doibase 10.1063/1.1699639} {\bibfield  {journal}
  {\bibinfo  {journal} {J. Appl. Phys.}\ }\textbf {\bibinfo
  {volume} {21}},\ \bibinfo {pages} {232} (\bibinfo {year} {1950})}\BibitemShut
  {NoStop}%
\bibitem [{\citenamefont {Thornburg}(2007)}]{AHFinder}%
  \BibitemOpen
  \bibfield  {author} {\bibinfo {author} {\bibfnamefont {J.}~\bibnamefont
  {Thornburg}},\ }\href {\doibase 10.12942/lrr-2007-3} {\bibfield
  {journal} {\bibinfo  {journal} {Living Rev. Relativity}\ }\textbf {\bibinfo
  {volume} {10}} (\bibinfo {year}
  {2007})}\BibitemShut {NoStop}%
\bibitem [{\citenamefont {Balay}\ \emph {et~al.}(1997)\citenamefont {Balay},
  \citenamefont {Gropp}, \citenamefont {McInnes},\ and\ \citenamefont
  {Smith}}]{petsc-efficient}%
  \BibitemOpen
  \bibfield  {author} {\bibinfo {author} {\bibfnamefont {S.}~\bibnamefont
  {Balay}}, \bibinfo {author} {\bibfnamefont {W.~D.}\ \bibnamefont {Gropp}},
  \bibinfo {author} {\bibfnamefont {L.~C.}\ \bibnamefont {McInnes}}, \ and\
  \bibinfo {author} {\bibfnamefont {B.~F.}\ \bibnamefont {Smith}},\ }in\
  \href@noop {} {\emph {\bibinfo {booktitle} {Modern Software Tools in
  Scientific Computing}}},\ \bibinfo {editor} {edited by\ \bibinfo {editor}
  {\bibfnamefont {E.}~\bibnamefont {Arge}}, \bibinfo {editor} {\bibfnamefont
  {A.~M.}\ \bibnamefont {Bruaset}}, \ and\ \bibinfo {editor} {\bibfnamefont
  {H.~P.}\ \bibnamefont {Langtangen}}}\ (\bibinfo  {publisher}
  {Birkh{\"{a}}user, Springer, New York},\ \bibinfo {year} {1997}),\ p.\
  \bibinfo {pages} {163}\BibitemShut {NoStop}%
\bibitem [{\citenamefont {Balay}\ \emph {et~al.}(2015)\citenamefont {Balay},
  \citenamefont {Abhyankar}, \citenamefont {Adams}, \citenamefont {Brown},
  \citenamefont {Brune}, \citenamefont {Buschelman}, \citenamefont {Dalcin},
  \citenamefont {Eijkhout}, \citenamefont {Gropp}, \citenamefont {Kaushik},
  \citenamefont {Knepley}, \citenamefont {McInnes}, \citenamefont {Rupp},
  \citenamefont {Smith}, \citenamefont {Zampini},\ and\ \citenamefont
  {Zhang}}]{petsc-user-ref}%
  \BibitemOpen
  \bibfield  {author} {\bibinfo {author} {\bibfnamefont {S.}~\bibnamefont
  {Balay}}, \bibinfo {author} {\bibfnamefont {S.}~\bibnamefont {Abhyankar}},
  \bibinfo {author} {\bibfnamefont {M.~F.}\ \bibnamefont {Adams}}, \bibinfo
  {author} {\bibfnamefont {J.}~\bibnamefont {Brown}}, \bibinfo {author}
  {\bibfnamefont {P.}~\bibnamefont {Brune}}, \bibinfo {author} {\bibfnamefont
  {K.}~\bibnamefont {Buschelman}}, \bibinfo {author} {\bibfnamefont
  {L.}~\bibnamefont {Dalcin}}, \bibinfo {author} {\bibfnamefont
  {V.}~\bibnamefont {Eijkhout}}, \bibinfo {author} {\bibfnamefont {W.~D.}\
  \bibnamefont {Gropp}}, \bibinfo {author} {\bibfnamefont {D.}~\bibnamefont
  {Kaushik}}, \bibinfo {author} {\bibfnamefont {M.~G.}\ \bibnamefont
  {Knepley}}, \bibinfo {author} {\bibfnamefont {L.~C.}\ \bibnamefont
  {McInnes}}, \bibinfo {author} {\bibfnamefont {K.}~\bibnamefont {Rupp}},
  \bibinfo {author} {\bibfnamefont {B.~F.}\ \bibnamefont {Smith}}, \bibinfo
  {author} {\bibfnamefont {S.}~\bibnamefont {Zampini}}, \ and\ \bibinfo
  {author} {\bibfnamefont {H.}~\bibnamefont {Zhang}},\ }\href
  {http://www.mcs.anl.gov/petsc/documentation/}{
  \bibinfo {institution} {Argonne National Laboratory}
  \bibinfo {type} {Technical Report}\ No. \bibinfo {number}
  {ANL-95/11---Revision 3.6},\ \bibinfo {year} {2015}} (unpublished)\BibitemShut {NoStop}%
\bibitem [{\citenamefont {Meijering}\ \emph {et~al.}(1999)\citenamefont
  {Meijering}, \citenamefont {Zuiderveld},\ and\ \citenamefont
  {Viergever}}]{Meijering99imagereconstruction}%
  \BibitemOpen
  \bibfield  {author} {\bibinfo {author} {\bibfnamefont {E.~H.~W.}\
  \bibnamefont {Meijering}}, \bibinfo {author} {\bibfnamefont {K.~J.}\
  \bibnamefont {Zuiderveld}}, \ and\ \bibinfo {author} {\bibfnamefont {M.~A.}\
  \bibnamefont {Viergever}},\ }\href {\doibase 10.1109/83.743854} {\bibfield  {journal} {\bibinfo
  {journal} {IEEE Trans. on Image Process.}\ }\textbf {\bibinfo
  {volume} {8}},\ \bibinfo {pages} {192} (\bibinfo {year} {1999})}\BibitemShut
  {NoStop}%
\bibitem [{\citenamefont {Dias}\ \emph {et~al.}(2014)\citenamefont {Dias},
  \citenamefont {Hartnett},\ and\ \citenamefont {Santos}}]{Dias:2014eua}%
  \BibitemOpen
  \bibfield  {author} {\bibinfo {author} {\bibfnamefont {O.~J.~C.}\
  \bibnamefont {Dias}}, \bibinfo {author} {\bibfnamefont {G.~S.}\ \bibnamefont
  {Hartnett}}, \ and\ \bibinfo {author} {\bibfnamefont {J.~E.}\ \bibnamefont
  {Santos}},\ }\href {\doibase 10.1088/0264-9381/31/24/245011} {\bibfield
  {journal} {\bibinfo  {journal} {Classical and Quantum Gravity}\ }\textbf
  {\bibinfo {volume} {31}},\ \bibinfo {pages} {245011} (\bibinfo {year}
  {2014})}\BibitemShut {NoStop}%
%%CITATION = ARXIV:1402.7047;%%
\bibitem [{\citenamefont {Figueras}\ \emph
  {et~al.}({\natexlab{b}})\citenamefont {Figueras}, \citenamefont {Kunesch},
  \citenamefont {Tunyasuvunakool},\ and\ \citenamefont
  {Wells}}]{DiffusionPaper}%
  \BibitemOpen
  \bibfield  {author} {\bibinfo {author} {\bibfnamefont {P.}~\bibnamefont
  {Figueras}}, \bibinfo {author} {\bibfnamefont {M.}~\bibnamefont {Kunesch}},
  \bibinfo {author} {\bibfnamefont {S.}~\bibnamefont {Tunyasuvunakool}}, \ and\
  \bibinfo {author} {\bibfnamefont {G.}~\bibnamefont {Wells}}\ }\href@noop {}
  {\bibfield  {journal} 
  (to be published)}\BibitemShut {NoStop}%
\end{thebibliography}%
\bibliographystyle{apsrev4-1}

\pagebreak

\makeatletter 
\renewcommand{\thefigure}{S\@arabic\c@figure}
\makeatother

\makeatletter 
\renewcommand{\theequation}{S\@arabic\c@equation}
\makeatother

\onecolumngrid

\begin{figure*}
{\large\bf Supplemental Material}
\end{figure*}

\twocolumngrid

\setcounter{figure}{0}
\setcounter{equation}{0}

%-------------------------------------------------------
% Diffusion
%-------------------------------------------------------
\PRLsection{Singularity diffusion}
Numerical relativity turns out to be considerably more challenging in non-standard settings such as higher dimensions
and/or at ultra relativistic speeds. In high-energy collisions in 4D, some lowering of the Courant factor was necessary
to improve the numerical stability \cite{Sperhake:2012me}. In higher dimensions, gauge parameters had to be finely tuned
\cite{Zilhao:2010sr,Witek:2013koa} or excision had to be used in conjunction with the standard puncture method \cite{Shibata:2010wz}.
An additional challenge in simulating black rings are the very steep gradients close to the singularity. These
gradients arise because the metric is very far from being conformally flat and the coordinate singularity cannot be fully
absorbed into the conformal factor $\chi$.
For rings this is particularly problematic because the singularity is extended,
in contrast with spherical black holes for which the singularity is point like.
In this section we outline how diffusion terms can be used to solve these problems in a very stable and simple way.

We found that the only problematic quantity is the conformal metric $\tilde \gamma_{ij}$ as it appears in the equations of motion with 2\textsuperscript{nd}
derivatives. While steep gradients arise in the extrinsic curvature and evolved conformal connection functions, they never caused problems in
our simulations. Inspired by artificial viscosity terms \cite{donea,lapidus,vonNeumann}
in computational fluid dynamics, we therefore add a term of the form
\begin{align} \label{eq:diffusion}
c_L \Delta x^2 g(\chi,|\partial \tilde \gamma_{ij}|) (\nabla^2 \tilde \gamma_{ij})^{\TF}
\end{align}
to the evolution equation of $\tilde \gamma_{ij}$, where $c_L$ is a dimensionless parameter, $\Delta x$ is the grid spacing
and $g(\chi,|\partial \tilde \gamma_{ij}|)$ is a sensing function which ensures that the diffusion term is only added
sufficiently inside the horizon and only when the gradients in $\tilde \gamma_{ij}$ become large.
Note that we enforce the diffusion term to be trace-free so that $\tilde \gamma = 1$ remains satisfied.

A diffusion term of the form \eqref{eq:diffusion} mostly affects small wavelength features. In particular, the timescale
over which a mode of wavelength $\lambda$ is diffused is approximately proportional to $(\lambda/\Delta x)^2$.
Note that, due to the inclusion of $\Delta x^2$ in \eqref{eq:diffusion}, this timescale automatically adapts when the
grid resolution is changed. With correctly calibrated $c_L$ and sensing function $g$, it is therefore possible to only diffuse modes
that could not be resolved.
Another advantage of the inclusion of $\Delta x^2$ is that the diffusion term does not impose a stricter Courant
condition than required in standard numerical relativity.

\begin{figure}
\centering
\begin{overpic}[scale=.125,unit=1mm]{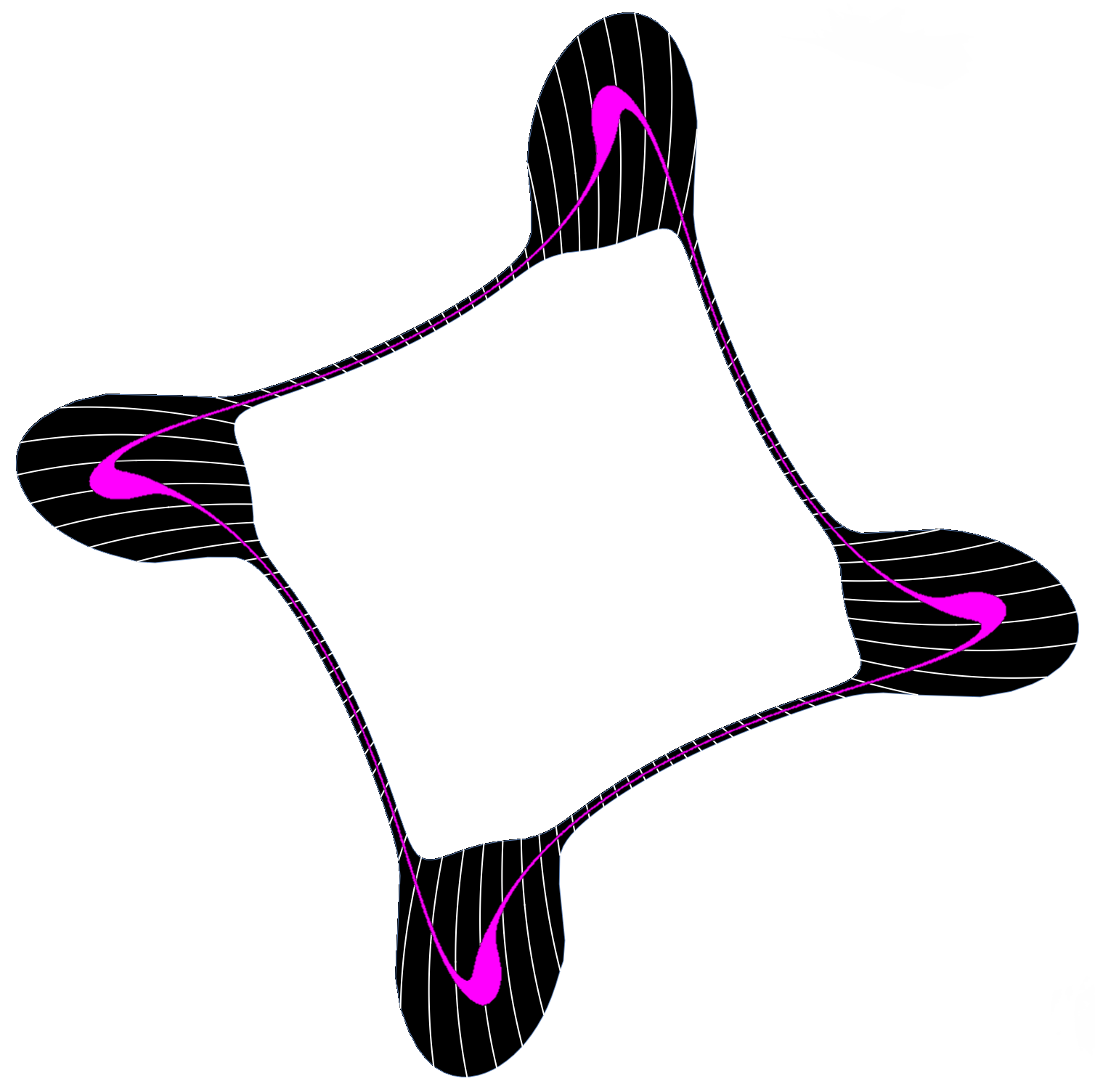}
\put(43.5,35.5){\frame{\includegraphics[scale=0.04]{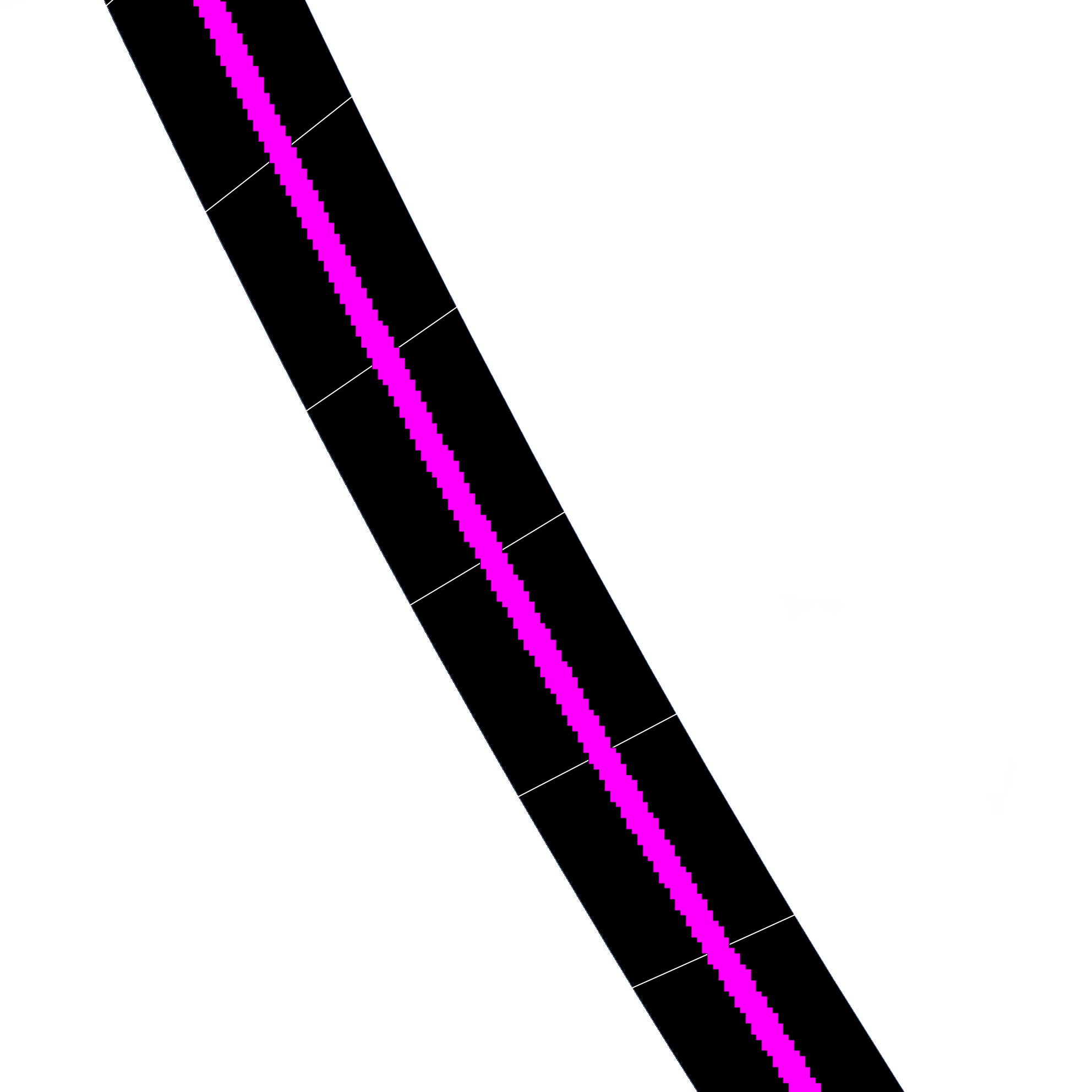}}}
\end{overpic}
\caption{Region of non-zero diffusion (pink) shown on top of the apparent horizon for $\nu=0.15$ with an $m=4$ mode, which has
become highly non-linear. The diffusion region covers much less than 50\% of the horizon's
interior, even at the thin necks.
\label{fig:diffusionHorizon}}
\end{figure}

In our simulations we chose $0.05 < c_L < 0.2$ and
\begin{align} \label{eq:g}
g(\chi, |\partial \tilde \gamma_{ij}|) = H(\chi_c-\chi) \sqrt{ \frac{2}{D(D-1)} \sum_{i,j,k} (\partial_k \tilde
\gamma_{ij})^2 },
\end{align}
where $H$ is the Heaviside step function, $D$ is the number of spacetime dimensions (five in our case)
and $\chi_c$ is a critical value which we chose between $0.015$ and $0.03$.
This cut-off for $\chi > \chi_c$ is used to ensure that the diffusion stays restricted to less than 50\% of the apparent
horizon. We used this method as it is hard to find apparent horizons reliably during the simulation, especially close to the topology
change.
However, we calculated the apparent horizons in the post-processing stage at every coarse level step, that is at a rate of
between $\tfrac{1}{4} \tilde R$ and $\tfrac{2}{3} \tilde R$, and checked that the diffusion term was indeed restricted
to less than 50\% of the horizon (see Fig.~\ref{fig:diffusionHorizon} for an example).
Since no information can escape the horizon and we have plenty of points across the apparent horizon, a diffusion term with this
configuration does not affect the results. We present specific tests for this in the last section of this document.

\begin{figure}
\centering
\begin{overpic}[scale=.11,unit=1mm]{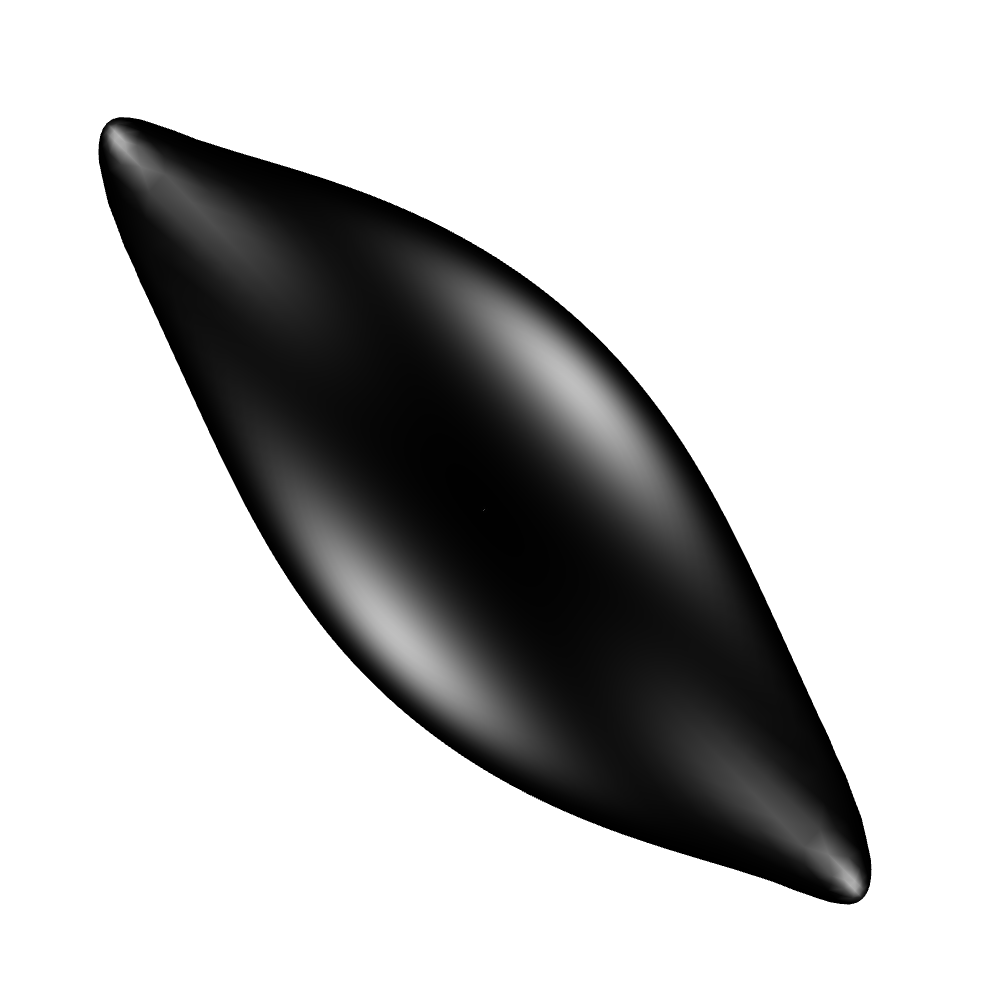}
  \put(17,1){(a)}
\end{overpic}
\begin{overpic}[scale=.15,unit=1mm]{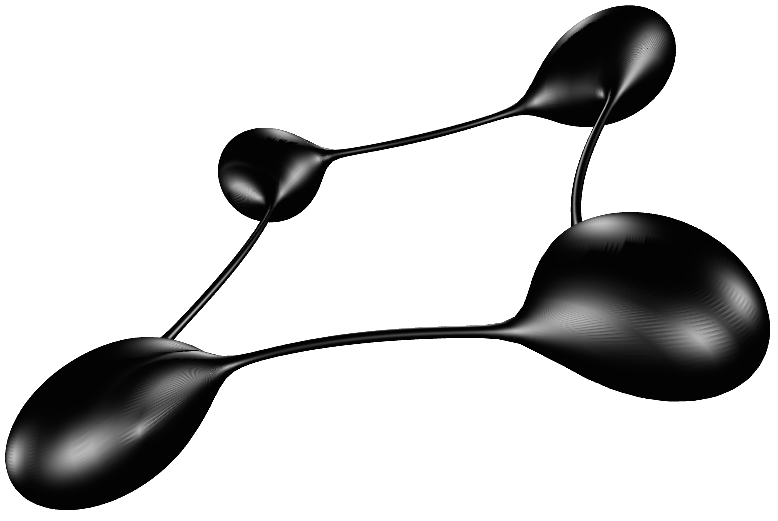}
  \put(17,1){(b)}
\end{overpic}
\begin{overpic}[scale=.15,unit=1mm]{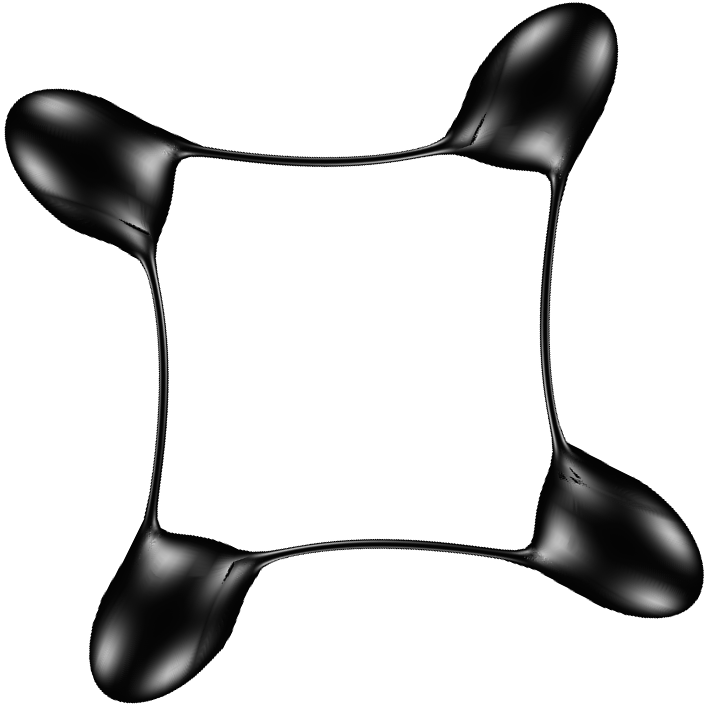}
  \put(20,1){(c)}
\end{overpic}
\begin{overpic}[scale=.0725,unit=1mm]{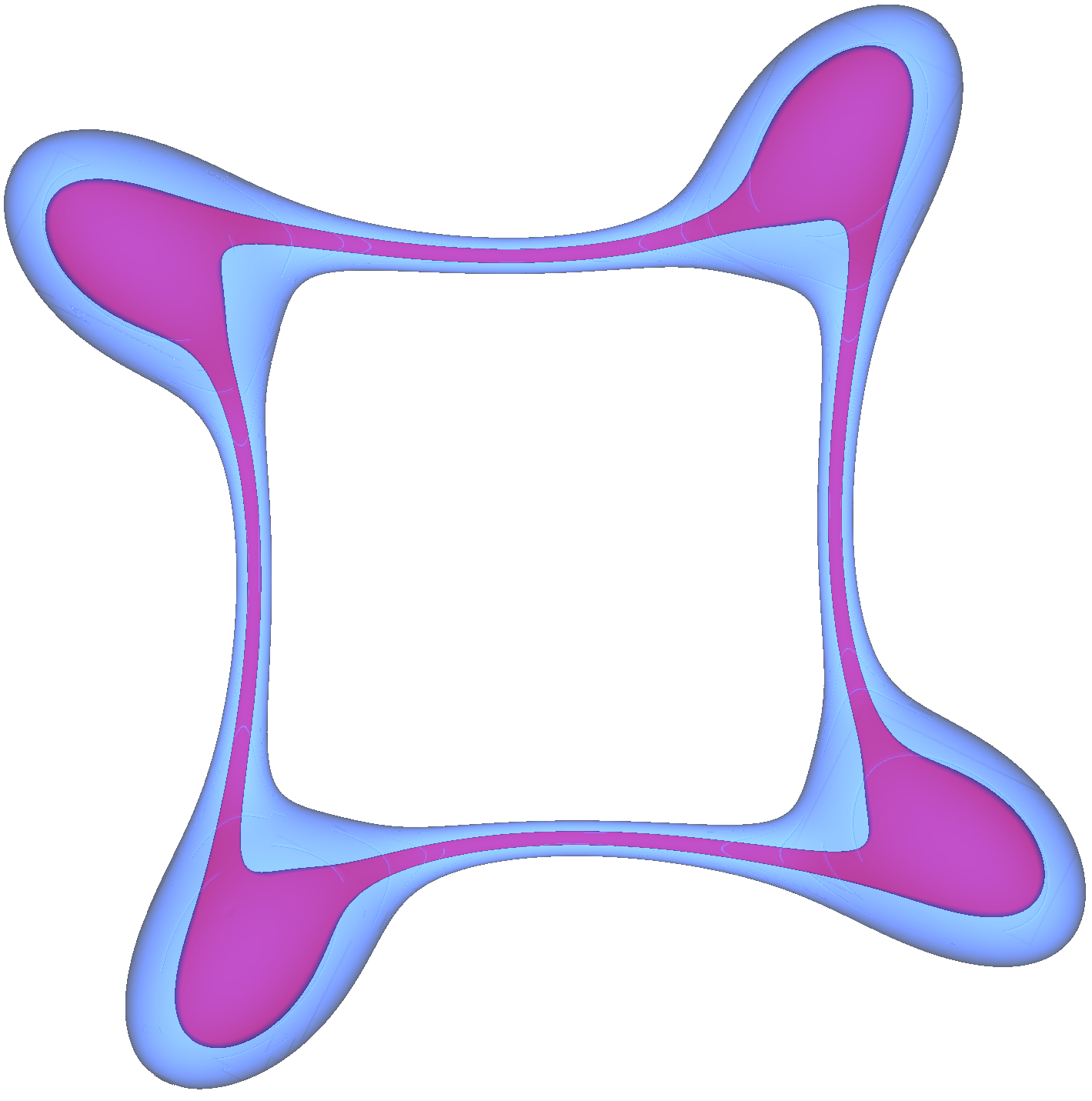}
  \put(19,1){(d)}
\end{overpic}
\caption{(a) First spherical apparent horizon that appears for the $\nu=0.2$ ring described in the main text. This settles down to a MP black hole. (b) and (c): two different views of the apparent horizon in the highly dynamical regime of the $\nu=0.15$ ring perturbed with a pure $m=4$ perturbation. The GL instability dominates, and big bulges connected by long and thin necks form during the evolution. (d) The $\chi=0.2$ (pink) and $\chi=0.4$ (light blue) contours for the same snapshot as in (c). The former contour approximates very well the apparent horizon along the thin necks, whilst the latter does so on the big bulges.}
\label{fig:AH_app}
\end{figure}

%-------------------------------------------------------
% Apparent horizon
%-------------------------------------------------------
\PRLsection{Apparent horizon for black rings}
In order to study the dynamics of the black hole's geometry, we need to locate the apparent horizon (AH) on each constant-$t$ slice $\Sigma$ in our numerical solutions. We adopt the usual definition of the AH as the outermost boundaryless surface $\mathcal{H}$ in $\Sigma$ on which outgoing null geodesics have zero expansion \cite{AHFinder}. More precisely, let $s^{i}$ be the normal of $\mathcal{H}$ in $\Sigma$, then the \emph{expansion}
\begin{equation}
\Theta \equiv \left( \gamma^{ij} - s^{i} s^{j} \right) \left( \nabla_{i} s_{j} - K_{ij} \right)
\label{eq:expansion}
\end{equation}
vanishes identically on $\mathcal{H}$, where $\gamma$ and $K$ are the induced metric and extrinsic curvature on $\Sigma$, respectively. In our setup, $\mathcal{H}$ inherits the $U(1)$ symmetry of the numerical solution and can therefore be treated as a 2-surface. If we assume that $\mathcal{H}$ can be parametrized by some coordinates $\left( \xi, \zeta \right)$, then we can construct a change of coordinates $(x,y,z) \mapsto (\xi,\zeta,\lambda)$ in the neighborhood of $\mathcal{H}$, such that $\mathcal{H}$ becomes the zero contour
$$\lambda - \Lambda(\xi,\zeta) \equiv 0,$$
for some unknown function $\Lambda(\xi,\zeta)$. The normal to $\mathcal{H}$ is then given by $s(\xi,\zeta) = \nabla(\lambda - \Lambda(\xi,\zeta))$, and equation (\ref{eq:expansion}) can now be viewed as an elliptic PDE $\Theta[\Lambda](\xi,\zeta) = 0$. For surfaces of $S^2$ (spherical) topology, we follow the usual practice in numerical relativity and use spherical polar coordinates $(\xi,\zeta,\lambda) \equiv (\theta,\phi,r)$, see Fig.~\ref*{fig:AH_app}(a) for an example. However, we also successfully adopted this approach for surfaces of $S^1 \times S^1$ (toroidal) topology. To do this, we closely follow the ring coordinates of \cite{Emparan:2001wn,Emparan:2006mm} and set $(\xi,\zeta,\lambda) \equiv (X,\psi,Y)$, where
\begin{equation}
\begin{gathered}
\psi = \tan^{-1}\mathopen{}\left(\mathclose{} y/x \right) \\
X = \frac{ R(\psi)^{2} - x^{2} - y^{2} - z^{2}}{\sqrt{\left( R(\psi)^{2} + x^{2} + y^{2} + z^{2} \right)^{2} - 4 \, R(\psi)^{2} \left( x^{2} + y^{2} \right)}} \\
Y = \frac{-R(\psi)^{2} - x^{2} - y^{2} - z^{2}}{\sqrt{\left( R(\psi)^{2} + x^{2} + y^{2} + z^{2} \right)^{2} - 4 \, R(\psi)^{2} \left( x^{2} + y^{2} \right)}} .
\end{gathered}
\label{eq:ringAH}
\end{equation}
Note that the usually-constant parameter $R$ is now allowed to vary with the angle $\psi$. This modification is necessary to cope with the highly stretched geometry of the thin rings, where the singularity is far from being a perfect circle in our working Cartesian coordinates. For each solution, we construct a suitable $R(\psi)$ by following small-valued contours of the conformal factor $\chi$. 

We discretize the resulting elliptic PDE using fourth-order finite differences and solve it using Newton line search method with automatic step size control as implemented in the PETSc library \cite{petsc-efficient,petsc-user-ref}. In order to interpolate geometric quantities from our AMR mesh onto $\mathcal{H}$, we implemented both fourth-order Lagrange polynomial and quintic convolution \cite{Meijering99imagereconstruction} interpolators. In theory, the latter has the advantage of producing an interpolant which is $C^3$ everywhere, at the expense of one lower convergence order in the second derivative. However, in practice, we found no significant differences between these two methods.

Our horizon finder is subject to the restriction that the set of points on the surface $\mathcal{H}$ must form a star-domain when projected onto the $(\xi,\zeta)$ plane. This is adequate for most of our results apart from later stages in the evolution of our two thinnest rings. The breakdown of the star-domain condition comes from two main sources: frame dragging and forming of large bulges. The first of these can be mitigated within our current approach by applying a `twisting' transformation in the opposite direction before applying (\ref{eq:ringAH}). The latter case will require a different, more flexible parametrization of $\mathcal{H}$. We leave the implementation of such a horizon finder for future work. However, we note that in this regime $\mathcal{H}$ seems to largely follow two contours of $\chi$: the thin necks roughly follow the $\chi \sim 0.2$ contour, whilst the bulges roughly follow the $\chi \sim 0.4$ contour. We therefore present the $\chi = 0.2$ contour as an indication of the qualitative appearance of $\mathcal{H}$ in lieu of a more flexible horizon finder. See Fig.~\ref{fig:AH_app} (c) and (d).

%-------------------------------------------------------
% Wave extraction
%-------------------------------------------------------
\PRLsection{Extraction of frequency and growth rate}
Since numerical relativity is an inherently non-linear tool, finding good estimates for frequency and growth rate of
the black ring instability in the linear regime is tricky, especially when there is competition between GL and elastic
mode. In particular, extracting a long linear signal requires
lowering the amplitude of the initial perturbation \eqref{eq:initdat}. However, due to grid noise one can only get
meaningful results for $A>10^{-6}$.

To obtain frequency and growth rate as well as the associated errors given a gravitational wave signal, we use the
following procedure: we first fit the entire wave signal of the linear regime to \eqref{eq:fit} to obtain a
rough estimate for the parameters. This estimate is then used to rescale the wave signal and fitting function
\eqref{eq:fit} such that the amplitude of the signal stays roughly constant. The effect of this is that subsequent
least-squares fits will not minimize the absolute errors but the errors measured as fraction of the amplitude at the
time. We found that this greatly improves the accuracy of the extracted parameters
when applied to fabricated data with Gaussian noise.

Finally, we fit the rescaled data in a window of length roughly $10$ units of computational
time as the window slides across the whole signal. Fig.~\ref{fig:waves} (bottom) shows the mean of the results; error
bars are given by the standard deviation. This last step allows us to quantify not only the uncertainty
of the fit but also the uncertainty due to the fact that our results are not obtained from a linear but a non-linear simulation, with
contamination from constraint violating modes towards the beginning and non-linearities towards the end. The resulting
uncertainty in the frequency is much lower than the uncertainty in the growth rate. The reason for this is that while
\eqref{eq:fit} fixes the frequency exactly, there is a slight degeneracy between the growth rates and the initial amplitudes
when performing the fit. The latter are unknown to us since we do not know how the amplitude of the
constraint-violating initial perturbation translates into initial amplitudes of the GL and the elastic mode.

\begin{figure}
\centering
\includegraphics[scale=0.7]{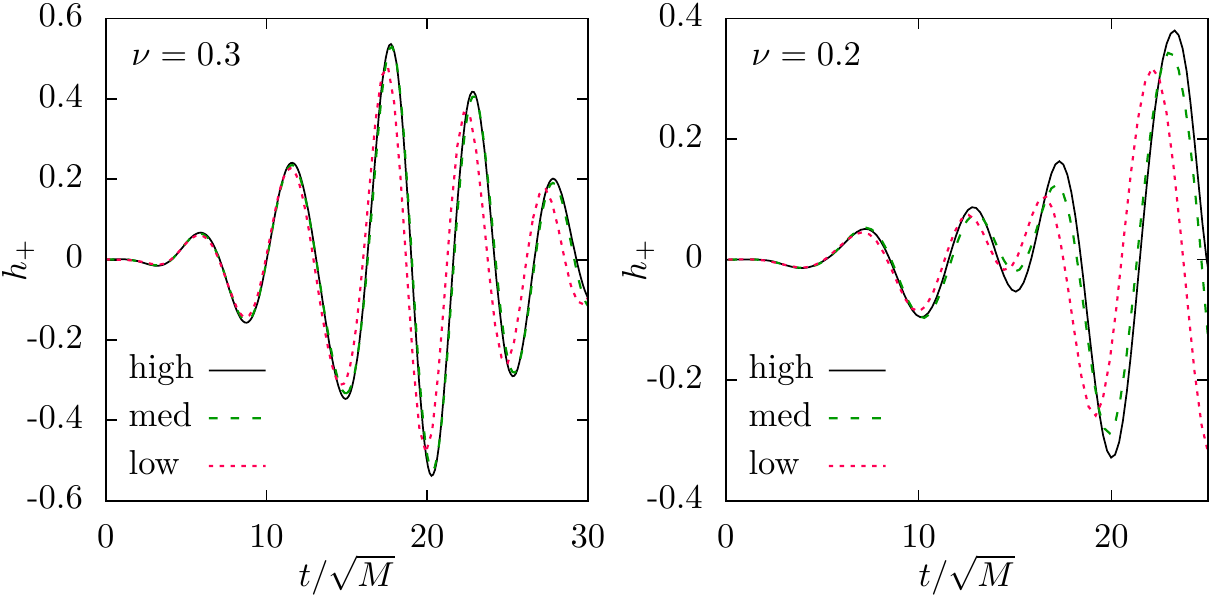}
\includegraphics[scale=0.7]{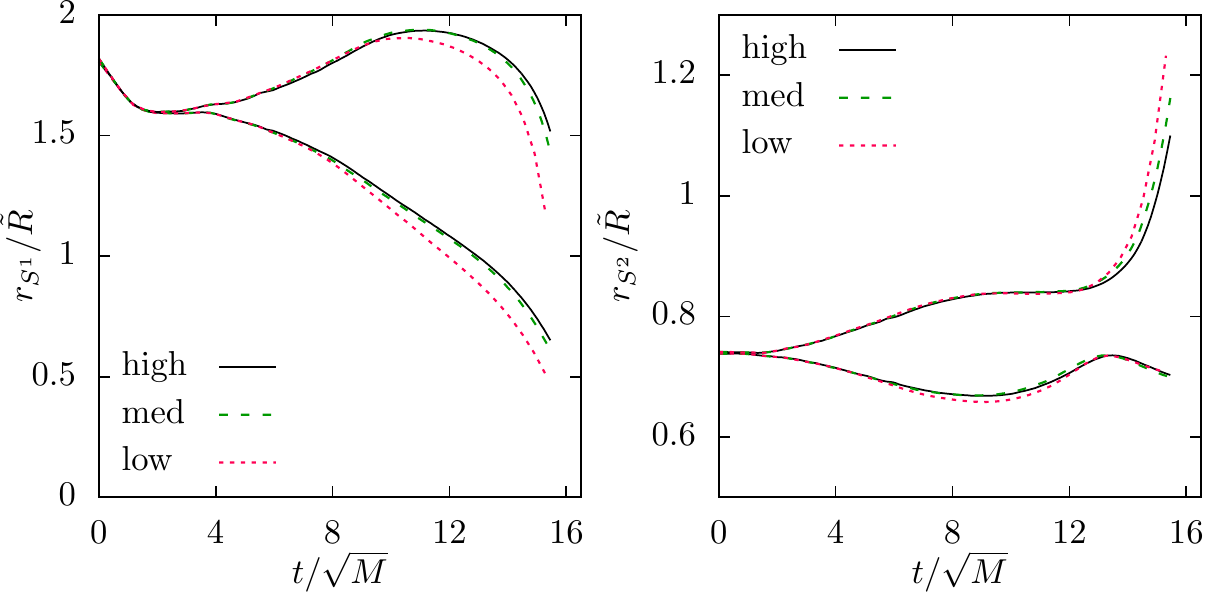}
\caption{\textit{Top:} Convergence tests of the wave data. The runs for $\nu=0.2$ are very expensive computationally and were
restricted to the regime before collapse to a spherical black hole. \textit{Bottom:} Convergence tests for the maximum
and minimum radius of the $S^1$ (left) and $S^2$ (right) for $\nu=0.3$. For all data there is a factor of $1.3$
between different resolutions. The highest resolution as measured on the finest refinement level is
$\Delta x = 0.003 \tilde R$ for $\nu=0.3$ and $\Delta x = 0.002 \tilde R$ for $\nu=0.2$. \label{fig:convergence}}
\end{figure}

%-------------------------------------------------------
% Numerical tests
%-------------------------------------------------------
\PRLsection{Numerical tests}
Here we present several numerical test to demonstrate the correctness of our results.
First, we present convergence tests for representative samples of the wave and horizon data presented in Fig.~\ref{fig:waves} and Fig.~\ref{fig:AH}. Fig.~\ref{fig:convergence} (\textit{top}) shows the wave
data for $\nu=0.2$ and $\nu=0.3$ for three different resolutions. Since the case $\nu = 0.2$ is very computationally
expensive, the resolution is slightly lower than for $\nu=0.3$ and we restrict the convergence test to the relevant regime
before collapse to a spherical black hole. The results clearly converge. Furthermore, the errors in frequency and growth rate
due to the finite resolution are much smaller than the errors due to the fit.
Fig.~\ref{fig:convergence} (\textit{bottom}) shows the minimum and maximum $S^1$ and $S^2$ radius for $\nu = 0.3$ for three
different resolutions. Again, the results clearly show convergence.
All the runs presented in the paper have resolution at least as high as
the medium resolution runs in Fig.~\ref{fig:convergence}, both in terms of number of points covering the horizon and
resolution in the wave zone.

Let us now turn to tests of the singularity diffusion inside the horizon.
First we note that the convergence tests above are also strong indicators that the diffusion term does not affect the
physical results. The reasons for this are twofold: firstly, as the resolution is increased the number of points between
the apparent horizon and the region where diffusion is applied increases. Secondly, the diffusion term
\eqref{eq:diffusion} is scaled with $\Delta x^2$ and therefore vanishes in the continuum limit.
As a final test we change the size of the diffusion region by varying the cut-off $\chi_c$ and monitor changes in the
wave signal. We expect the wave signal to converge as $\chi_c$ is reduced.
Table \ref{tab:diffCutoffTest} shows the results for frequency and growth rate of the initial stage of the instability
and the final ring-down after the ring has collapsed to a spherical black hole.
The results clearly show convergence and indicate that for the values of $\chi_c$ used for this paper ($0.015 < \chi_c <
0.03$) the effect of the diffusion term is negligible.

\begin{table}
Growth: 
\vspace*{0.15cm} \\
\begin{tabular}{c|c|c|c|c|c} 
$\chi_c$ & 0.05 & 0.03 & 0.018 & $\Delta(0.05,0.018)$ & $\Delta(0.03,0.018)$ \\
\hline
$\Re \varpi_{el}$ & 0.736 & 0.764 & 0.751 & 2.0\% & 1.7\% \\
$\Im \varpi_{el}$ & 0.162 & 0.172 & 0.169 & 4.2\% & 1.6\% \\
$\Re \varpi_{GL}$ & 1.21 & 1.26 & 1.24 & 2.6\% & 2.0\% \\
$\Im \varpi_{GL}$ & 0.083 & 0.093 & 0.092 & 9.2\% & 1.4\% \\
\end{tabular}\\
\vspace*{0.5cm}
Ring-down: 
\vspace*{0.15cm} \\
\begin{tabular}{c|c|c|c|c|c} 
$\chi_c$ & 0.05 & 0.03 & 0.018 & $\Delta(0.05,0.018)$ & $\Delta(0.03,0.018)$ \\
\hline
$\Re \varpi$ & 1.19 & 1.22 & 1.21 & 1.6\% & 1.2\% \\
$\Im \varpi$ & -0.1657 & -0.1681 & -0.1679 & 1.3\% & 0.11\%\\
\end{tabular}
\caption{Results for frequency ($\Re \varpi$) and growth rate ($\Im \varpi$) for different $\chi_c$
for the initial growth of the instability and the final ring-down to a MP black hole for a $\nu=0.3$ ring.
For the growth $\varpi = \omega/(2\pi T)$, where $T$ is the
temperature of the unperturbed ring and for the ring-down $\varpi = \omega \sqrt{M_{\mathrm{MP}}}$, where $M_{\mathrm{MP}}$
is the mass of the final MP black hole.
The modulus of the difference between the results for two different values of $\chi_c$ is denoted by
$\Delta(\chi_c^1,\chi_c^2)$.
\label{tab:diffCutoffTest}}
\end{table}

Finally we remark that for those simulations which end up in collapse into a spherical black hole, once the system has settled, we can compute the area of the apparent horizon and the circumference on the rotation plane of the final black hole. This allows us to estimate the mass and angular momentum of the resulting MP black hole. Extracting the quasinormal modes from the wave signal we can compare with existing results in the literature \cite{Dias:2014eua}, and we find good agreement. 

\end{document}